\documentclass[12pt]{elsart}
\usepackage{feynmp,epsfig,graphics}
\def\be{\begin{eqnarray}}
\def\ee{\end{eqnarray}}
\def\bc{\begin{center}}
\def\ec{\end{center}}

\newcommand{\tr}{\rm tr \,}

\def\Im{{\rm Im\,}}

\begin{document}
GSI-Preprint-2003-19
\begin{frontmatter}
\title{On meson resonances and chiral symmetry }
\author[GSI,TU]{M.F.M. Lutz}
\author[NBI]{and E.E. Kolomeitsev}
\address[GSI]{Gesellschaft f\"ur Schwerionenforschung (GSI),\\
Planck Str. 1, 64291 Darmstadt, Germany}
\address[TU]{Institut f\"ur Kernphysik, TU Darmstadt\\
D-64289 Darmstadt, Germany}
\address[NBI]{The Niels Bohr Institute\\ Blegdamsvej 17, DK-2100 Copenhagen, Denmark}
\begin{abstract}
We study meson resonances with quantum numbers $J^P=1^+$ in terms
of the chiral SU(3) Lagrangian. At leading order a parameter-free
prediction is obtained for the scattering of Goldstone bosons off
vector mesons with $J^P=1^-$ once we insist on approximate
crossing symmetry of the unitarized scattering amplitude. A resonance
spectrum arises that is remarkably close to the empirical
pattern. In particular, we find that the strangeness-zero resonances
$h_1(1380)$, $f_1(1285)$ and $b_1(1235)$ are formed due to strong
$K\,\bar K_\mu$ and $\bar K\,K_\mu$ channels. This leads to large coupling constants
of those resonances to the latter states.

\end{abstract}
\end{frontmatter}

\section{Introduction}

In recent works \cite{LK00,LK01,LK02,Granada,Copenhagen} it was demonstrated that chiral SU(3) symmetry
predicts parameter-free $J^P\!=\!\frac{1}{2}^-$ and $J^P\!=\!\frac{3}{2}^-$
baryon resonances. It was observed that the resonance states turn into
bound states in the heavy SU(3) limit with $m_{\pi,K,\eta}\simeq 500$ MeV
but disappear altogether in the light SU(3) limit with
$m_{\pi,K,\eta}\simeq 140$ MeV. Related works \cite{grnpi,grkl,Oset-prl,Oset-plb,Jido03} that
did not insist on approximate crossing symmetric scattering amplitudes \cite{LK02} and
therefore involve a larger set of free parameters adjusted to the data are in qualitative
agreement with
the parameter-free computation \cite{Granada} for the s-wave $J^P\!=\!\frac{1}{2}^-$ baryon
resonances. Even earlier in the 60s a series of works \cite{Wyld,Dalitz,Ball,Rajasekaran,Wyld2}
predicted a wealth of s-wave baryon resonance generated by coupled-channel dynamics.
Those works were based on a SU(3)-symmetric interaction Lagrangian that is closely related
to the leading order chiral Lagrangian. All these results strongly support a conjecture put
forward by the authors that baryon resonances that do not belong to the
large-$N_c$ ground state of QCD are generated by coupled-channel dynamics \cite{LK01,LK02,LH02}.
If this conjecture is further substantiated one would expect an analogous
mechanism to be at work in the meson sector, i.e. we conjecture in this
work that meson resonances not belonging to the large-$N_c$
ground state of QCD are generated by coupled-channel dynamics as well.
In contrast to the baryon sector it is not so immediate which meson states
one should identify to be the large-$N_c$ ground states of QCD. Without
controversy we would take for the latter the SU(3) Goldstone bosons,
($\pi(140), K(494), \eta(545)$),
with $J^P\!=\!0^-$ and the lightest vector mesons
($\rho_\mu(770), \omega_\mu(782), K_\mu(892), \phi_\mu(1020)$) with
$J^P\!=\!1^-$. However, it is unclear whether to include also the
lightest meson resonances with quantum numbers $J^P\!=\!0^+$ and
$J^P\!=\!1^+$. From recent coupled-channel analyses that were based on
the chiral SU(3) Lagrangian and were able to quantitatively describe the
$\pi \pi$ phase shifts in the $J^P\!=\!0^+$ sector we would conclude that
meson resonances with scalar quantum numbers should be generated dynamically.
A prime example is the $f_0(980)$ resonance \cite{Rupp86,WI90,JPHS95,OOP99,NVA02,NP02} which long ago was found to be a
$K \bar K-$molecule. This is analogous to the finding that the parity
partners of the large-$N_c$ baryon ground states, i.e. baryon resonances
with $J^P\!=\!\frac{1}{2}^-$ and $J^P\!=\!\frac{3}{2}^-$, are predicted
by chiral coupled-channel dynamics. In view of these results it is quite
natural to also describe the axial-vector meson resonances in terms of
coupled-channel dynamics rather than identifying the latter to be part of the
large-$N_c$ ground state of QCD. Our point of view differs here from that one expressed
in recent works \cite{PPR01,deRafael01} where the axial-vector mesons are part of a minimal hadronic
ansatz of large-$N_c$ QCD.

In this work we study the scattering of Goldstone bosons off vector mesons
using the leading order chiral Lagrangian. Our results are parameter free
once we insist on approximate crossing symmetric scattering amplitudes.
We find that chiral symmetry predicts the existence of the axial-vector
meson resonances ($h_1(1170), h_1(1380), f_1(1285), a_1(1260), b_1(1235),
K_1(1270), K_1(1400)$) with a spectrum that is surprisingly close to the
empirical one. A result analogous
to the baryon sector \cite{LK02,Granada,Copenhagen} is found: in the heavy SU(3) limit with
$m_{\pi, K, \eta} \simeq 500$ MeV and $m_{\rho, \omega, K^*, \phi}\simeq 900$
MeV the resonance states turn into bound states forming two degenerate octets and
one singlet of the SU(3) flavor group with masses  1360 MeV and 1280 MeV respectively.
Taking the light SU(3) limit with $m_{\pi, K, \eta} \simeq 140$ MeV
and $m_{\rho, \omega, K^*, \phi}\simeq 700$ MeV we do not observe any
bound-state nor resonance signals anymore. Since the leading order interaction
kernel scales with $N_c^{-1}$ the resonances disappear also
in large-$N_c$ limit. Using physical masses a
pattern arises that compares surprisingly well with the empirical properties of
the $J^P\!=\!1^-$ meson resonances.

\section{Chiral coupled-channel dynamics: the $\chi$-BS(3) approach}

The starting point to study the scattering of Goldstone bosons off
vector mesons is the chiral SU(3) Lagrangian. Though it is straight
forward to construct the infinite tower of covariant interaction terms \cite{Krause}
the inclusion of massive vector mesons into the chiral Lagrangian
is non-trivial: it requires a careful check whether
chiral power counting rules can be realized manifestly. A solution to this problem
is  a systematic heavy-meson expansion \cite{Jenkins:Manohar:Wise:95}.
An alternative solution is offered by the minimal chiral $\overline{MS}_\chi$-scheme \cite{LK02}
developed in the meson-baryon sector recently (see also \cite{BL99,FGJS03,FSGS03}). The latter
is based on dimensional regularization as applied to Feynman
diagrams derived from the relativistic chiral Lagrangian and therefore
preserves also all Ward identities if applied in perturbation theory. It exploits an
ambiguity of how to introduce a subtraction scheme in dimensional
regularization that arises once the theory is expected to be applicable
only around a heavy mass scale. Certain algebraic consistency  identities
that probe the full relativistic loop functions outside the validity domain
of the theory but that are respected by the $\overline{MS}$ scheme are given up
with the benefit that chiral power counting rules can be implemented
manifestly. Consistency is achieved by subtracting not only the standard
poles at $d=4$ but in addition poles at $d=3$ \cite{PDS98} that arise in the chiral
limit. The resulting loop functions are finite and well defined and
comply with the leading chiral moment predicted by chiral power counting
rules. The loop functions may be expanded further identifying subleading
chiral moments. It is obvious that the $\overline{MS}_\chi$ scheme is applicable
to the chiral Lagrangian involving any heavy fermion or boson field.

We proceed and identify the leading order Weinberg-Tomozawa interaction Lagrangian
density \cite{Wein-Tomo,Krause,Ecker89}
\begin{eqnarray}
{\mathcal L}_{WT}(x) &=&
-\frac{1}{16\,f^2}\,\tr \Big( [\Phi^\mu(x)\,, (\partial^\nu \Phi_\mu(x) )]_-
\,[\phi (x) , (\partial_\nu\,\phi(x))]_-
\Big) \,,
\label{WT-term}
\end{eqnarray}
describing the interaction of the Goldstone bosons field $\phi$ with
a massive vector-meson field $\Phi_\mu$. The parameter
$f$ in (\ref{WT-term}) is known from the weak decay process of the
pions. We use $f= 90$ MeV through out this work.
In (\ref{WT-term}) we omit additional terms that do not contribute to the
on-shell scattering amplitude of Goldstone bosons off vector-meson at
tree-level. Such terms are suppressed and not probed in a leading order calculation
using the $\chi-$BS(3) scheme. Similarly we do not anticipate a dependence on the choice
of the realization of the spin 1 field, though this issue may deserve further
studies once sub-leading terms are included. Due to the on-shell reduction scheme
discussed in detail in the following section we do not expect different results if we
apply the tensor realization rather than the vector realization used in
(\ref{WT-term}). The equivalence of the two realizations in some cases was
demonstrated in \cite{Ecker89} to leading orders.

Since we will assume perfect
isospin symmetry it is convenient to decompose
the fields into their isospin multiplet
\begin{eqnarray}
&& \phi = \tau \cdot \pi (140)
+ \alpha^\dagger \cdot  K (494) +  K^\dagger(494) \cdot \alpha
+ \eta(547)\,\lambda_8\,,
\nonumber\\
&& \Phi_\mu = \tau \cdot \rho_\mu(770)
+ \alpha^\dagger \cdot  K_\mu(892) +  K_\mu^\dagger(892) \cdot \alpha
\nonumber\\
&& \qquad +
\Big( {\textstyle{2\over 3}}+ {\textstyle{1\over \sqrt{3}}} \,\lambda_8  \Big)\, \omega_\mu(782)
+\Big(  {\textstyle{\sqrt{2}\over 3}}-\sqrt{{\textstyle{2\over 3}}}\,\lambda_8\Big)\,
\phi_\mu(1020)  \,,
\nonumber\\
&& \alpha^\dagger = {\textstyle{1\over \sqrt{2}}}\,(\lambda_4+i\,\lambda_5 ,\lambda_6+i\,\lambda_7 )\,,\qquad
\tau = (\lambda_1,\lambda_2, \lambda_3)\,,
\label{}
\end{eqnarray}
with for instance $K_\mu =(K_\mu^{(+)},K^{(0)}_{\mu})^t$ and
$\rho_\mu =(\rho_\mu^{(1)},\rho_\mu^{(2)},\rho_\mu^{(3)} )$. The matrices
$\lambda_i$ are the standard Gell-Mann generators of the SU(3) algebra.
The numbers
in the brackets recall the approximate masses of the fields in units of MeV.

As was emphasized in \cite{LK02} the chiral SU(3) Lagrangian should not be
used in perturbation theory except at energies sufficiently below all thresholds. Though the
infinite set of irreducible diagrams can be successfully approximated by the standard
perturbative chiral expansion that is no longer true for the infinite set of
reducible diagrams once the energies are sufficiently large to support hadronic
scattering processes. Whereas the former diagrams are controlled by the typical
small parameter $m_K/(4\pi f)$ the latter ones probe a parameter of unit size
$m_K^2/(8 \pi f^2)$, which invalidates any perturbative expansion. From an
effective field theory point of view it is mandatory to sum the reducible
diagrams. This is naturally achieved by considering the Bethe-Salpeter
scattering equation,
\begin{eqnarray}
&& T_{\mu \nu}(\bar k, k; w) =
K_{\mu \nu}(\bar k, k; w) +\int
\frac{d^4l}{(2\,\pi)^4}\,K_{\mu \beta}(\bar k, l; w)\,
G^{\alpha \beta}(l;w)\,
T_{\alpha \nu}(l, k; w) \,,
\nonumber\\
&& G_{\mu \nu}(l;w)=\frac{i}{ ({\textstyle{1\over 2}}\,w-l)^2-m^2+i\,\epsilon}\,
\frac{g_{\mu\nu}- ({\textstyle{1\over 2}}\,w+l)_\mu\,({\textstyle{1\over 2}}\,w+l)_\nu/M^2}{({\textstyle{1\over 2}}\,w+l)^2-M^2+i\,\epsilon}\,,
\label{B-eq}
\end{eqnarray}
where we suppress the coupled-channel structure for simplicity.
We use physical values for the Goldstone boson and vector meson
masses $m$ and $M$ respectively. Since the intermediate vector mesons
have in part a substantial decay width we allow for spectral distributions of the
broadest vector mesons, the $\rho_\mu$- and $K_\mu$-mesons. In channels involving
the $\rho_\mu-$ or $K_\mu-$meson the two-particle propagator,
$G$, in (\ref{B-eq})
is folded with spectral functions obtained at the one-loop level describing the
decay processes $\rho_\mu \to \pi \,\pi$ and $K_\mu \to \pi\, K$.

\begin{table}
\tabcolsep=-.1mm
\begin{tabular}{||c|c|c|c||}
\hline\hline
$(0,-2)$ &
$(1,-2)$ &
$(\frac12,-1)$ &
$(\frac32, -1)$
\\\hline
$({\textstyle{1\over \sqrt{2}}}\,\overline{K}^t\,i\,\sigma_2 \overline{K}_\mu)$ &
$({\textstyle{1\over \sqrt{2}}}\,\overline{K}^t\,i\,\sigma_2 \,\vec \sigma\,\overline{K}_\mu )$ &
$\left(\begin{array}{c}
{\textstyle{1\over \sqrt{3}}}\,\pi\cdot
\sigma \,i\,\sigma_2\, \overline{K}^t_\mu) \\
({\textstyle{1\over \sqrt{3}}}\, \rho_\mu \cdot \sigma\, i\,\sigma_2 \,\overline{K} )\\
 (i\,\sigma_2\, \overline{K}^t\, \omega_\mu)\\
 (\eta \, i\,\sigma_2\, \overline{K}^t_\mu)\\
(i\,\sigma_2\, \overline{K}^t\, \phi_\mu )
\end{array}\right)$ &
$\left(\begin{array}{c}
(\pi\cdot T \,i\,\sigma_2\,\overline{K}^t_\mu)\\
(\rho_\mu \cdot T\,i\,\sigma_2\, \overline{K}^t )
 \end{array}\right)$
\\\hline\hline
$(0,+2)$ &
$(1,+2)$ &
$(\frac12,+1)$ &
$(\frac32, +1)$
\\\hline
$( {\textstyle{1\over \sqrt{2}}}\,K^t\,i\,\sigma_2 K_\mu)$ &
$( {\textstyle{1\over \sqrt{2}}}\,K^t\,i\,\sigma_2 \,\vec \sigma\,K_\mu )$ &
$\left(\begin{array}{c}
({\textstyle{1\over \sqrt{3}}}\,\pi\cdot \sigma \, K_\mu)\\
({\textstyle{1\over \sqrt{3}}}\, \rho_\mu \cdot \sigma\,K )\\
(K\, \omega_\mu)\\
(\eta \,K_\mu)\\
( K\, \phi_\mu)
\end{array}\right)$ &
$\left(\begin{array}{c}
(\pi\cdot T \,K_\mu)\\
(\rho_\mu \cdot T\,K\, )
\end{array}\right)$
\\\hline\hline
$(0^+,0)$ &
 $(0^-,0)$ &
 $(1^+,0)$  &
 $(1^-,0)$
 \\ \hline
$\left({\textstyle{1\over 2}}\,( \overline{K}\,
K_\mu-\overline{K}_\mu\,K) \right)$ &
$\left(\!\!\begin{array}{c}
({\textstyle{1\over \sqrt{3}}}\, \pi\cdot \rho_\mu) \\
(\eta \, \omega_\mu)\\
{\textstyle{1\over 2}}\,( \overline{K}\,
K_\mu+\overline{K}_\mu\,K) \\
(\eta\, \phi_\mu)
\end{array}\!\!\right)$&
$\left(\!\!\begin{array}{c}
(\pi\, \omega_\mu)\\
(\pi\, \phi_\mu)\\
(\eta \, \rho_\mu) \\
{\textstyle{1\over 2}}\,( \overline{K}\,\sigma\,
K_\mu+\overline{K}_\mu\,\sigma\,K)
\end{array}\!\!\right)$ &
$\left(\!\!\begin{array}{c}
({\textstyle{1\over {i\sqrt{2}}}}\, \pi \times \rho_\mu) \\
({\textstyle{1\over 2}}\,(
\overline{K}\,\sigma\, K_\mu-\,\overline{K}_\mu \,\sigma \,K)
\end{array}\!\!\right)$
\\
\hline\hline
\multicolumn{4}{||c|}{$(2,0)$}
\\ \hline
\multicolumn{4}{||c|}{${\textstyle{1\over 2}}(\pi_i \,\rho^\mu_j
+\pi_j \,\rho^\mu_i) -{\textstyle{1\over 3}}\, \delta_{ij}\,\pi \cdot
\rho^\mu $ }
 \\\hline\hline
\end{tabular}
\caption{The column $R^{(I^G,S)}_\mu(q,p)$ for isospin ($I$), G-parity ($G$)
and strangeness ($S$). The Pauli matrices
$\sigma_i$ act on isospin doublet fields $K,\,K_\mu$ and $\bar K,\, \bar K_\mu$.
The 4$\times$2 matrices $T_j$  describe the
transition from isospin-$\frac{1}{2}$ to $\frac{3}{2}$ states. We use the normalization
implied by $\vec T \cdot \vec T^\dagger =1 $ and
$T^\dag_i T_j=\delta_{ij}-{\textstyle{1\over 3}}\,\sigma_i\,\sigma_j $.}
\label{tab:states}
\end{table}

The Bethe-Salpeter interaction kernel $K_{\mu \nu}(\bar k,k;w)$ is the sum
of all two-particle irreducible diagrams, i.e. at leading order it reads
\begin{eqnarray}
&& K^{\mu \nu}(\bar k, k;w) = -\frac{C_{WT}}{4\,f^2}\, (p+\bar p)\cdot
(q+\bar q) \,g^{\mu \nu}\,,
\nonumber\\
&& w= p+q= \bar p+ \bar q \,,\qquad k = {\textstyle{1\over 2}}\,(p-q)
\,,\qquad \bar k = {\textstyle{1\over 2}}\,(\bar p-\bar q)\,,
\label{K-int}
\end{eqnarray}
where the coupled-channel structure is suppressed for convenience. It is
straight forward to restore the latter giving the coefficient $C_{WT}$
a matrix structure. The scattering problem decouples into thirteen orthogonal
channels specified by isospin ($I$), G-parity ($G$) and strangeness ($S$) quantum numbers.
This decomposition is implied by
a corresponding decomposition of the Weinberg-Tomozawa interaction Lagrangian
density represented in momentum space \cite{LK02}
\begin{eqnarray}
&&{\mathcal L}_{WT}(\bar k,k;w)= \sum_{I^G,S}\,R^{(I^G,\,S),\dagger}_\mu(\bar q, \bar p)\,
g^{\mu \alpha}\,
K^{(I^G,S)}_{\alpha \beta}(\bar k, k;w)\,
g^{\beta \nu}\,R_\nu^{(I^G,\,S)}(q,p)\,,
\nonumber\\
&& (I^G,S)=
((0,\pm 2), (1,\pm 2), ({\textstyle{1\over 2}}, \pm 1), ({\textstyle{3\over 2}}, \pm 1),
(0^\pm, 0), (1^\pm,0), (2,0))\,,
\label{lwt}
\end{eqnarray}
where the objects $R_\nu^{(I^G,\,S)}(q,p)$ are specified in Tab. \ref{tab:states} for all
possible channels.

\begin{table}
\tabcolsep=2.4mm
\begin{center}
\begin{tabular}{||c||p{9mm}|p{9mm}|p{9mm}|p{9mm}|p{9mm}|p{9mm}|p{9mm}|p{9mm}|p{9mm}||}
\hline
($I^G, S$)    & ($0, \pm 2$)  & ($1, \pm 2$) & ($\frac{1}{2}, \pm 1$) &
($\frac{3}{2}, \pm 1$) & ($0^+,0$) & ($0^-,0$) &($1^+,0$) &($1^-,0$) & ($2,0$)  \\
 \hline\hline
11 & $\phantom{-}0$ & $-2$ & $\phantom{-}2$& $-1$ & $\phantom{-}3$ & $\phantom{-}4$ & $\phantom{-}0$ & $\phantom{-}2$ & $-2$  \\ 
  \hline
12 &-- &-- &$\phantom{-}\frac{1}{2}$ & $-1$ &-- & $\phantom{-}0$ & $\phantom{-}0$ & $\phantom{-}\sqrt{2}$ &--  \\ 
  \hline
22 &-- &-- & $\phantom{-}2$  &$-1$ & -- &$\phantom{-}0$ & $\phantom{-}0$ &$\phantom{-}1$ &-- \\ 
  \hline
13 &-- & -- &$-\frac{\sqrt{3}}{2}$ &-- &-- & $\phantom{-}\sqrt{3}$ &$\phantom{-}0$ &-- &-- \\ 
  \hline
23 &-- & -- &$\phantom{-}0$ &-- &-- & $\phantom{-}\sqrt{3}$ &$\phantom{-}0$ &-- &-- \\ 
  \hline
33 &-- &-- &$\phantom{-}0$ &-- &-- &$\phantom{-}3$ &$\phantom{-}0$ &-- &--  \\ 
  \hline
14 &-- & -- &$\phantom{-}0$ &-- &-- & $\phantom{-}0$ &$\phantom{-}1$ &-- &--  \\ 
  \hline
24 &-- &-- &$-\frac{3}{2}$ &-- &-- & $\phantom{-}0$ & $-\sqrt{2}$ &-- &--  \\ 
  \hline
34 &-- & -- & $-\frac{\sqrt{3}}{2}$ &-- & -- & $-\sqrt{6}$ &$\phantom{-}\sqrt{3}$&-- &--  \\ 
  \hline
44 &-- &-- & $\phantom{-}0$  &-- &-- &$\phantom{-}0$ &$\phantom{-}1$&-- &-- \\ 
  \hline
15 & -- &-- & $\phantom{-}\sqrt{\frac{3}{2}}$ &-- &-- &-- &-- &-- &-- \\ 
   \hline
25 & --&-- & $\phantom{-}0$  &-- &-- &-- &-- &-- &--  \\ 
  \hline
35 & -- &-- & $\phantom{-}0$  &-- &-- &-- &-- &-- &-- \\ 
  \hline
45 & -- &-- & $\phantom{-}\sqrt{\frac{3}{2}}$  &-- &-- &-- &-- &-- &-- \\ 
   \hline
55 & -- &-- & $\phantom{-}0$  &-- &-- &-- &-- &-- &--  \\ 
\hline
\end{tabular}
\caption{The coefficients $C^{(I,S)}$ of the Weinberg-Tomozawa
term that characterize the  interaction of Goldstone bosons with vector mesons
as introduced in (\ref{K-int}, \ref{lwt}).} \label{tab:coeff}
\end{center}
\end{table}

From a field theoretic point of view once
the interaction kernel and the two-particle propagator are specified
the Bethe-Salpeter equation (\ref{B-eq}) determines the scattering
amplitude $T_{\mu \nu}(\bar k, k;w)$. However,
in order to arrive at a scattering amplitude that  does not depend on the
choice of interpolating fields at given order in a truncation
of the scattering kernel it is necessary to perform an on-shell
reduction. In \cite{LH02} it was suggested to introduce the latter with respect
to the unique set of covariant projectors that solve the Bethe-Salpeter equation.

\subsection{On-shell reduction scheme}

The scattering amplitude
$T_{\mu \nu}(\bar k, k;w)$ as it is determined by the Bethe-Salpeter equation
does not have a well defined off-shell extrapolation.
The latter depends explicitly on the choice of the chiral Lagrangian. However, the solution
of the scattering equation requires the scattering amplitude for off-shell kinematics.
The question arises how do we ever arrive at any meaning-full result. A related issue
is the evaluation of higher n-point Green's functions. The latter ones require
necessarily the knowledge of the off-shell part of the two-body amplitude
for a given choice of fields. Any systematic scheme should specify
not only the on-shell two-body amplitude but also the form of higher n-point functions.
Thus it is important to derive the off-shell part of the two-body amplitude in a given scheme.

The idea put forward in \cite{LK02} exploits covariance as a tool to construct a minimal
off-shell extrapolation of the scattering amplitude as to render the Bethe-Salpeter scattering
equation well defined within dimensional regularization. Here we derive the off-shell part of
the two-body amplitude for any given choice of fields.
The on-shell part of the scattering amplitude,
\begin{eqnarray}
&& T_{\mu \nu}(\bar k,k;w) =  T^{\rm on-shell}_{\mu \nu}(\bar q,q;w)
 + T^{\rm off-shell}_{\mu \nu}(\bar q,q;w)\,,
\nonumber\\
&&  T^{\rm on-shell}_{\mu \nu}(\bar q,q;w)= \sum_{J,P,a,b}\,M^{(J P)}_{ab}(\sqrt{s})\,
{\mathcal Y}^{(J P)}_{\mu\nu, ab}(\bar q,q;w)\,,
\label{t-exp}
\end{eqnarray}
is expanded in a series of projectors
${\mathcal Y}^{(J P)}_{\mu \nu ,ab}(\bar q,q;w)$, defined for any off-shell kinematics,
and invariant amplitudes that depend  on
$\sqrt{s}$ only. Any projector ${\mathcal Y}^{(J P)}_{\mu \nu ,ab}(\bar q,q;w)$ is
characterized by its total angular momentum $J$ and parity $P$ quantum number. If
in a given channel $J^P$ a degeneracy is left due to the coupling of various
helicity states the projectors acquire an additional matrix structure.
A crucial property of the projectors ${\mathcal Y}^{(J P)}_{\mu \nu ,ab}(\bar q,q;w)$ is
their regularity. The presence of kinematical singularities in the projectors
would lead to a pathological behavior when inserting those
into the Bethe-Salpeter equation and trying to establish the frame-independence of the
scattering amplitude. A detailed derivation of the projectors is given in the Appendix.

We recall and further elaborate on the on-shell reduction scheme suggested in \cite{LK02}.
For a given choice of interpolating fields the full off-shell
scattering amplitude may require $T^{\rm off-shell}_{\mu \nu} \neq 0 $ in (\ref{t-exp}), however,
we argue that the terms proportional to the invariant amplitudes
must always be present independent of the choice of fields.
An effective interaction kernel $V_{\mu \nu}(\bar q,q;w)$ is introduced such that
if feeded into the Bethe-Salpeter equation it produces the on-shell scattering amplitude
$T^{\rm on-shell}_{\mu \nu}$, i.e. in functional notation
\begin{eqnarray}
T^{\rm on-shell} = V +V \cdot G \cdot T^{\rm on-shell} \,,
\label{BS-on-shell}
\end{eqnarray}
where $G$ is the two-particle propagator.
It is instructive to work out the structure of the off-shell part of the
scattering amplitude in this scheme. Straight forward manipulations lead to
\begin{eqnarray}
T^{\rm off-shell} = \Big( \big(1-V \cdot G\big)\cdot  \big( K-V\big)^{-1}\cdot \big( 1-G\cdot V\big) -G\Big)^{-1}\,.
\label{t-off-1}
\end{eqnarray}
The result (\ref{t-off-1}) is useful as a starting point for further developments but
as it stands it is not very instructive. It does not manifestly show the off-shell
nature of the amplitude and  also it does not suggest how to consistently expand the amplitude
for a given choice of interpolating fields. Progress is made by introducing three off-shell
interaction kernels $V_L,V_R$ and $V_{LR}$ where $V_R$ ($V_L$) vanishes if the
initial (final) particles are on-shell. The interaction kernel $V_{LR}$ is defined to
vanish if evaluated with either initial or final particles on-shell. The latter objects
are defined by:
\begin{eqnarray}
K &=& V+
(1-V \cdot G)\cdot V_{L}+ V_{R}\cdot (1-G\cdot V)\,
\nonumber\\
&+& (1-V\cdot G)\cdot V_{LR}\cdot (1-G\cdot V)
- V_R\cdot \frac{1}{1-G\cdot V_{LR}}\cdot G \cdot V_L \,.
\label{K-decomp}
\end{eqnarray}
The decomposition of the Bethe-Salpeter interaction kernel is unique and
can be applied to an arbitrary interaction kernel
once it is defined what is meant with the 'on-shell' part of any two-particle amplitude.
The latter we define as the part of the amplitude that has a decomposition
into the set of projectors introduced in (\ref{t-exp}). It is clear that performing a
chiral expansion of $K$ and $V$ to some order $Q^n$ leads to a straight forward
identification of the off-shell kernels $V_L,V_R$ and $V_{LR}$ to the same
accuracy. The particular way the off-shell interaction was introduced guarantees
consistency of the scheme. This is demonstrated by the exact result
\begin{eqnarray}
&& T^{\rm off-shell} =
\Big( V_L\cdot \frac{1}{1-G\cdot V}+V_{LR}\Big)\cdot \Big(1-G\cdot \bar V \Big)^{-1}
\nonumber\\
&& \qquad  -  \Big( V_L\cdot \frac{1}{1-G\cdot V}+V_{LR}\Big)\cdot \Big(1-G\cdot \bar V \Big)^{-1}
 \cdot G\,
 \Big( \frac{1}{1-V\cdot G}\,V_R+V_{LR}\Big)
\nonumber\\
&& \qquad
- V_{LR}
+ \Big(1-\bar V\cdot G \Big)^{-1} \Big( \frac{1}{1-V\cdot G}\,V_R+V_{LR}\Big)
 \,,
\nonumber\\
&& \bar V = V_{LR}+ V_L\cdot \frac{1}{1- G\cdot V } + \frac{1}{1- V\cdot G }\cdot V_R
\nonumber\\
&& \qquad -\frac{1}{1-V\cdot G}\cdot V_R\cdot \frac{1}{1-G\cdot V_{LR}}\cdot G\cdot V_L\cdot
 \frac{1}{1-G\cdot V}\,,
\label{t-off-2}
\end{eqnarray}
which proves that the off-shell amplitude vanishes if evaluated with on-shell kinematics.
The result (\ref{t-off-2}) suggests a systematic expansion of the off-shell part of the
scattering amplitude. The unitarization of the on-shell amplitude requires to count
$V\cdot G \sim Q^0$. Since any off-shell kernel meeting the two-particle
propagator $G$ does not generate a unitarity cut by construction, standard chiral counting rules
should be applied for the objects $G\cdot V_{L}\,, G\cdot V_{LR} $ and
$V_{R}\cdot G\,,  V_{LR}\cdot G $. Thus a unitary chiral expansion of the off-shell amplitude is
induced by an expansion thereof in powers of the off-shell kernels $V_L,V_R$ and $V_{LR}$.
At leading order we find
\begin{eqnarray}
&& T^{\rm off-shell} = V_L \cdot \frac{1}{1-G\cdot V}+
\frac{1}{1-V\cdot G}\cdot V_R + V_{LR} + \cdots \,,
\label{}
\end{eqnarray}
illustrating that the off-shell part of the amplitude requires necessarily a
summation once a summation is used for the on-shell amplitude.
This is an important result since it suggests a systematic way how to evaluate
higher n-point Green functions in a unitary chiral expansion scheme.
The latter requires necessarily the knowledge of the off-shell
part of the two-body amplitude for a given choice of fields \footnote{To arrive at
finite results for the off-shell amplitude may need in some cases additional counter
terms of the chiral Lagrangian that are not probed in the renormalization of the on-shell
part of the scattering amplitude (see e.g. \cite{Appelquist81}).}.

\subsection{Renormalization scheme and crossing symmetry}

Unlike in standard chiral perturbation theory the renormalization
of a unitarized chiral perturbation theory is non-trivial and therefore
requires particular care. Due to the defining properties of the projectors,
${\mathcal Y}_{\mu \nu}^{(J P)}(\bar q, q;w)$,
and of the effective interaction kernel, $V_{\mu \nu}(\bar q,q;w)$, the latter can be
expanded into a series of the former,
\begin{eqnarray}
V_{\mu \nu}(\bar k,k;w) = \sum_{J, P,a,b}\,V^{(J P)}_{ab}(\sqrt{s})\,
{\mathcal Y}^{(J P)}_{\mu \nu,ab}(\bar q,q;w) \,.
\label{v-exp}
\end{eqnarray}
The coefficient functions $V^{(J P)}_{ab}(\sqrt{s})$ are evaluated in chiral perturbation
theory and therefore standard renormalization schemes are applicable. The on-shell part of
the scattering amplitude takes the simple form,
\begin{eqnarray}
&& T^{\rm on-shell}_{\mu \nu}(\bar k ,k ;w )  = \sum_{J,P}\,M^{(J P)}(\sqrt{s}\,)\,
{\mathcal Y}^{(J P)}_{\mu \nu}(\bar q, q;w) \,,
\nonumber\\
&& M^{(J P)}(\sqrt{s}\,) = \Big[ 1- V^{(J P)}(\sqrt{s}\,)\,J^{(J P)}(\sqrt{s}\,)\Big]^{-1}\,
V^{(J P)}(\sqrt{s}\,)\,,
\label{}
\end{eqnarray}
with a set of divergent loop functions $J^{(J P)}(\sqrt{s}\,)$. The crucial issue is how to
renormalize the loop functions. In \cite{LK00,LK01,LK02} it was suggested to introduce a
physical scheme defined by the renormalization condition,
\begin{eqnarray}
T_{\mu \nu}^{(J P)}(\bar k,k;w)\Big|_{\sqrt{s}= \mu } =
V_{\mu \nu}^{(J P)}(\bar k,k;w)\Big|_{\sqrt{s}= \mu } \,,
\label{ren-cond}
\end{eqnarray}
where the subtraction scale $\mu = \mu(I,S)$ depends on isospin and strangeness but
is independent on $J^P$. It was argued in \cite{LK02} that the optimal choice of the
subtraction point can be determined by the requirement that the scattering amplitude
is approximatively crossing symmetric. Moreover it was demonstrated that the renormalization
condition (\ref{ren-cond}) is complete, i.e. the condition (\ref{ren-cond})
suffices to render the scattering amplitude finite. Before discussing in some detail the
choice of the subtraction points $\mu (I,S)$ let us elaborate on the structure of the
loop functions. The merit of our scheme is that dimensional regularization can be used
to evaluate the latter ones. Here we exploit the results (\ref{y-minus}, \ref{y-plus}) that any
given projector is a finite polynomial in the available 4-momenta. This implies that the
loop functions can be expressed in terms of a log-divergent
master function, $I(\sqrt{s}\,)$, and reduced tadpole terms,
\begin{eqnarray}
&& J^{(J P)}(\sqrt{s}\,)= N^{(J P)}(\sqrt{s}\,)\,\Big(I(\sqrt{s}\,) -I(\mu )\Big) \,,
\nonumber\\
&& I(\sqrt{s}\,)=\frac{1}{16\,\pi^2}
\left( \frac{p_{cm}}{\sqrt{s}}\,
\left( \ln \left(1-\frac{s-2\,p_{cm}\,\sqrt{s}}{m^2+M^2} \right)
-\ln \left(1-\frac{s+2\,p_{cm}\sqrt{s}}{m^2+M^2} \right)\right)
\right.
\nonumber\\
&&\qquad \qquad + \left.
\left(\frac{1}{2}\,\frac{m^2+M^2}{m^2-M^2}
-\frac{m^2-M^2}{2\,s}
\right)
\,\ln \left( \frac{m^2}{M^2}\right) +1 \right)+I(0)\;,
\label{i-def}
\end{eqnarray}
where $\sqrt{s}= \sqrt{M^2+p_{cm}^2}+ \sqrt{m^2+p_{cm}^2}$. The normalization factor
$N^{(J P)}(\sqrt{s}\,)$ is a polynomial in $\sqrt{s}$ and the mass parameters. In
(\ref{i-def}) the renormalization scale dependence of the
scaler loop function $I(\sqrt{s}\,)$ was traded in favor of a dependence on a subtraction point
$\mu$. The loop functions $J^{(J,P)}(\sqrt{s}\,)$ are consistent with chiral
counting rules only if the subtraction scale $\mu \simeq M$ is chosen close to the 'heavy'
meson mass \cite{LK00,LK01,LK02}. Furthermore we dropped additional terms that are proportional to
reduced tadpole contributions. The latter ones are real and must be moved into the
effective interaction kernel in order to arrive at the
decoupling of projectors with different quantum numbers \cite{LK02}. Since
tadpole contributions show in general a polynomial $\sqrt{s}$ dependence the
renormalization condition (\ref{ren-cond}) would not suffice to render the
loop function finite in the presence of such contributions.
In \cite{LK02} it was shown that keeping reduced tadpole terms in the loop functions
leads to a renormalization of s-channel exchange terms that is in conflict
with chiral counting rules. We emphasize that the projectors have the
important property that in the case of broad intermediate states the implied
loop functions follow from (\ref{i-def}) by a simple folding with the spectral
distributions of the two intermediate states.

Using the results of the Appendix (\ref{non-uni-def:a},\ref{non-uni-def}) the
normalization factors $N^{(J P)}(\sqrt{s}\,)$ in (\ref{i-def}) are readily derived
\begin{eqnarray}
&& N_{ab}^{(J P_+)}(\sqrt{s}\,) =
\left(
\begin{array}{cc}
\frac{p_{\rm cm}^{2\,(J-1)}}{(J+1)}\,\Big( (2\,J+1) + J\,\frac{p_{\rm cm}^2}{M^2} \Big) &
\;\;\;\sqrt{\frac{J}{J+1}}\,\frac{p_{\rm cm}^{2\,J}\,\sqrt{M^2+p_{\rm cm}^2}}{M^2} \\
\sqrt{\frac{J}{J+1}}\,\frac{p_{\rm cm}^{2\,J}\,\sqrt{M^2+p_{\rm cm}^2}}{M^2} & \frac{p_{\rm cm}^{2\,(J+1)}}{M^2}
\end{array}
\right)_{ab}\,,
\nonumber\\
&& N^{(J P_-)}(\sqrt{s}\,) = p_{\rm cm}^{2\,J} \,, \qquad P_+ =(-1)^{J+1}\,,\qquad P_-=(-1)^J\,,
\label{}
\end{eqnarray}
where we point out that the loop functions $J^{(J,+)}(\sqrt{s}\,)$
acquire off-diagonal elements. This is a consequence of a non-unitary transformation applied to
the helicity states (\ref{non-uni-def}). As demonstrated in the Appendix covariant projectors can
only be introduced with respect to states that are not orthogonal. The
threshold behavior of the normalization factor $N^{(J P)}(\sqrt{s}\,)$ associated with a given
projector tells the leading angular momentum ($L$) characteristic. For instance the projectors
${\mathcal Y}^{(J P_-)}$ and ${\mathcal Y}^{(J P_+)}_{11}$ carry $L= J$ and $L=J-1$
respectively. It is important, however, to realize that the coupled-channel projectors are not
defined with respect to states of good angular momentum $L$.

The renormalization condition (\ref{ren-cond}) reflects the basic assumption our effective
field theory is built on, namely, that at subthreshold energies the scattering amplitudes may
be evaluated in standard chiral perturbation theory with the typical expansion parameter
$m_K/(4 \,\pi f) < 1 $ with $f \simeq 90$ MeV. Once the available energy is sufficiently high
to permit elastic two-body scattering a further typical dimensionless parameter
$m_K^2/(8\,\pi f^2) \sim 1$ arises. Since this ratio is uniquely linked to the presence of a
two-particle unitarity cut it is sufficient to sum those contributions keeping the perturbative
expansion of all terms that do not
develop a two-particle unitarity cut. This is achieved by (\ref{BS-on-shell}, \ref{K-decomp}, \ref{t-off-2}).
In order to recover the perturbative nature of the subthreshold scattering amplitude
the subtraction scale $ M-m < \mu < M+m$ must be chosen in between the s- and u-channel
elastic unitarity branch points \cite{LK02}. In \cite{LK02}
it was suggested that s-channel and u-channel unitarized amplitudes should be glued together at
subthreshold kinematics. A smooth result is guaranteed if the full amplitudes match the
interaction kernel $V$ close to
the subtraction scale $\mu$ as imposed by (\ref{ren-cond}). In this case the crossing symmetry
of the interaction kernel, which follows directly from its perturbative evaluation,
carries over to an approximate crossing symmetry of the
full scattering amplitude. This construction reflects our basic assumption that diagrams
showing an s-channel or u-channel unitarity cut need to be summed to all orders only at energies
where the diagrams develop their imaginary part.

The reader should be reminded that at energies below its
u-channel unitarity cuts a partial-wave amplitude can be reconstructed uniquely in
terms of the scattering amplitudes of its crossed reaction. In this case the
crossed amplitudes are probed at energies above their s-channel unitarity thresholds only.
Thus, our final partial-wave amplitudes properly glued together at subthreshold energies respect
crossing symmetry exactly at energies above the s-channel and below the u-channel
unitarity cuts by construction. At subthreshold energies in between the s- and u-channel cuts
an approximate crossing symmetry is guaranteed by the matching condition (\ref{ren-cond}).
In cases like $\pi \,\omega_\mu$ scattering the crossed channel is redundant, in the sense
that all observable quantities can be expressed in terms of the direct channel only.

Stringent consistency condition for the optimal subtraction scales are derived by considering
photo-reactions like $\gamma \,\rho_\mu$  scattering. Since this system is coupled via
$\gamma \,\rho \to \pi \rho$ to the hadronic process we study here, one may include the
$\gamma \,\rho$ as a state part of the $S=0$ coupled-channel system. In this case the matching
of the s- and u-channel iterated amplitudes requires $\mu(I,0) = M_{\rho(770)}$ identically.
Similarly the subtraction scale, $\mu(I,\pm 1) = M_{K(892)}$  follows upon considering the
$\gamma \,K_\mu$ reactions. In the $S=\pm 2$ sector we use the same subtraction
scale as in the $S=0$ sector since the two sectors are related by a crossing transformation,
i.e. the $K\,K_\mu$ and $\bar K\,K_\mu$ amplitudes are transformed into each other by exchanging
$K \leftrightarrow \bar K$. We should mention a slight ambiguity. In the $S=0$ sector
we could have argued in terms of $\gamma \,\omega$ or $\gamma \,\phi$ reactions rather than
$\gamma \,\rho$. Since the three vector mesons $\rho_\mu, \omega_\mu$ and $\phi_\mu$ are
mass degenerate in the large-$N_c$ limit of QCD this ambiguity is of subleading importance.

Given the subtraction scales as derived above the leading-order calculation is parameter
free. Of course chiral correction terms lead to further so
far unknown parameters  which need to be adjusted to data.
Within the $\chi-$BS(3)
approach such correction terms enter the effective interaction kernel $V$ rather than leading to
a change of the subtraction scales. In particular the leading correction effects
are determined by the counter terms of chiral order $Q^2$.
The effect of altering the subtraction scales away from their optimal values
can be compensated for by incorporating counter terms in the chiral Lagrangian that carry
order $Q^3$. Our scheme  has the advantage that once the
parameters describing subleading effects are determined in a subset of sectors one has
immediate predictions for all sectors ($I,S$). In order to estimate the size of correction
terms one may vary the subtraction scales around their optimal values.

\section{Results}

\begin{figure}[t]
\begin{center}
\includegraphics[width=14.0cm,clip=true]{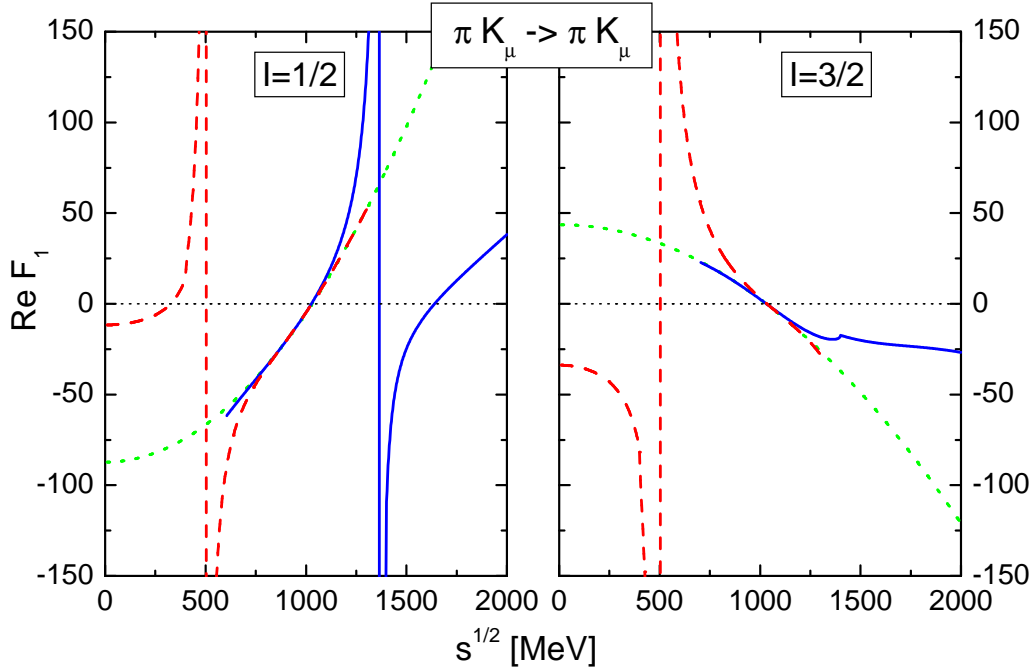}
\end{center}
\caption{Isospin $\frac{1}{2}$ (left panel) and $\frac{3}{2}$ (right panel)
forward scattering amplitude, $F_1(q,q;w)=F_1(\sqrt{s})$ (see (\ref{t-decomp})),
describing the $\pi \,K_\mu$ scattering process
in the 'heavy' SU(3) limit. Sharp vector meson masses are used. The
panels show three lines, the line extending to the right (left) shows the s-channel (u-channel)
unitarized scattering amplitude. The dotted lines represent the amplitude evaluated at
tree-level.} \label{fig:0}
\end{figure}

We present out results on s-wave scattering of Goldstone bosons off vector mesons using
the leading order chiral SU(3) Lagrangian. Meson resonances with quantum number (I,S) and
$J^P\!=\!1^+$ manifest themselves as poles in the corresponding scattering amplitudes
$T^{(I,S)}_{\mu \nu}(\bar q,q;w)$. We will suppress the index $J^P$ in the following
studying exclusively the $J^P\!=\!1^+$ sector.
The scattering amplitude takes the form
\begin{eqnarray}
&& T^{(I,S)}_{\mu \nu}(\bar q,q;w) = \sum_{ab}\, {\mathcal Y}_{ab, \mu \nu}(\bar q,q;w)\,M^{(I,S)}_{ab}(\sqrt{s}\,)
+\cdots \,,
\nonumber\\
&&  M^{(I,S)}(\sqrt{s}\,) = \Big[ 1- V^{(I,S)}(\sqrt{s}\,)\,J^{(I,S)}(\sqrt{s}\,)\Big]^{-1}\,
V^{(I,S)}(\sqrt{s}\,)\,,
\label{final-t}
\end{eqnarray}
where we suppressed further contribution
to the scattering amplitude $T^{(I,S)}_{\mu \nu}$ that are off-shell or not of s-wave type.
The required projectors in (\ref{final-t}) follow from (\ref{y-plus}) with $J=1$,
\begin{eqnarray}
&&  {\mathcal Y}_{11, \mu \nu} = \frac{3}{2}\,\frac{w_\mu\,w_\nu}{w^2}-  \frac{3}{2}\, g_{\mu \nu} \,,
\quad  {\mathcal Y}_{12, \mu \nu}  = - \frac{3}{\sqrt{2}}\,
\Big( q_\mu- \frac{q\cdot w}{w^2}\,w_\mu\Big)\,\frac{w_\nu}{\sqrt{w^2}}
\nonumber\\
&& {\mathcal Y}_{22, \mu \nu}  = 3\,
\Big( \frac{(\bar q\cdot w)\,(w \cdot q)}{w^2}-(\bar q\cdot q)\Big)\,\frac{w_\mu\,w_\nu}{w^2} \,,\qquad
\nonumber\\
&&  {\mathcal Y}_{21, \mu \nu}  = - \frac{3}{\sqrt{2}}\,
\frac{w_\mu}{\sqrt{w^2}}\,\Big( \bar q_\nu- \frac{\bar q\cdot w}{w^2}\,w_\nu\Big)\,.
\label{}
\end{eqnarray}
The invariant scattering amplitudes are determined by the effective interaction
kernel $V^{(I,S)}(\sqrt{s}\,)$ and the loop functions $J^{(I,S)}(\sqrt{s}\,)$. In the
$S= 2$ sector which involves only a single channel, $K\,K_\mu$, the matrix of
loop functions takes the form,
\begin{eqnarray}
J_{ab}^{(I,S)}(\sqrt{s}\,) =
\left(
\begin{array}{cc}
\frac{3}{2} + \frac{p_{\rm cm}^2}{2\,M^2}  &
\;\;\;\frac{p_{\rm cm}^{2}\,\sqrt{M^2+p_{\rm cm}^2}}{\sqrt{2}\,M^2} \\
\frac{p_{\rm cm}^{2}\,\sqrt{M^2+p_{\rm cm}^2}}{\sqrt{2}\,M^2} & \frac{p_{\rm cm}^{4}}{M^2}
\end{array}
\right)_{ab}\!\Big(I(\sqrt{s}\,)-I(\mu(I,S)) \Big)\,,
\label{}
\end{eqnarray}
with  $M=M_{K(892)}$ and $m=m_{K(494)}$. In the general
case the matrix of loop functions acquires additional dimensions reflecting the
presence of inelastic channels. The scalar loop function
$I(\sqrt{s}\,)$ was given in (\ref{i-def}). For the optimal subtraction scales $\mu(I,S)$
we obtained,
\begin{eqnarray}
&& \mu(I,0) =  M_{\rho(770)}\,,\qquad
\mu(I,\pm 1) = M_{K(892)}\,,\qquad
\mu(I,\pm 2) = M_{\rho(770)} \,.
\label{}
\end{eqnarray}
It remains to provide explicit expressions for the effective interaction kernel
$V^{(I,S)}(\sqrt{s}\,)$. The leading-order chiral SU(3) Lagrangian (\ref{WT-term}) implies
\begin{eqnarray}
V_{11}^{(I,S)}(\sqrt{s}\,) &=& \frac{C^{(I,S)}_{WT}}{12\,f^2}\, \Big(
3\,s-M^2-\bar M^2-m^2-\bar m^2
-\frac{M^2-m^2}{s}\,(\bar M^2-\bar m^2)\Big) \,,
\nonumber\\
V_{22}^{(I,S)}(\sqrt{s}\,) &=& -\frac{C^{(I,S)}_{WT}}{12\,f^2}\,,
\qquad V_{12}^{(I,S)}(\sqrt{s}\,)=V_{21}^{(I,S)}(\sqrt{s}\,)=0\,,
\label{}
\end{eqnarray}
where $(m,M)$ and $(\bar m, \bar M)$ are the masses of initial and final mesons. The matrix
of coefficients $C^{(I,S)}_{WT}$ is given in Tab. \ref{tab:coeff}.
It should be pointed out that though the leading order form of $V_{11}^{(1,+)}$ is determined
by the Weinberg-Tomozawa interaction term that this is not the case for $V_{22}^{(1,+)}$.
Therefore it is legitimate to use $V_{22}^{(1,+)}=0$ here.

\begin{figure}[t]
\begin{center}
\includegraphics[width=14.0cm,clip=true]{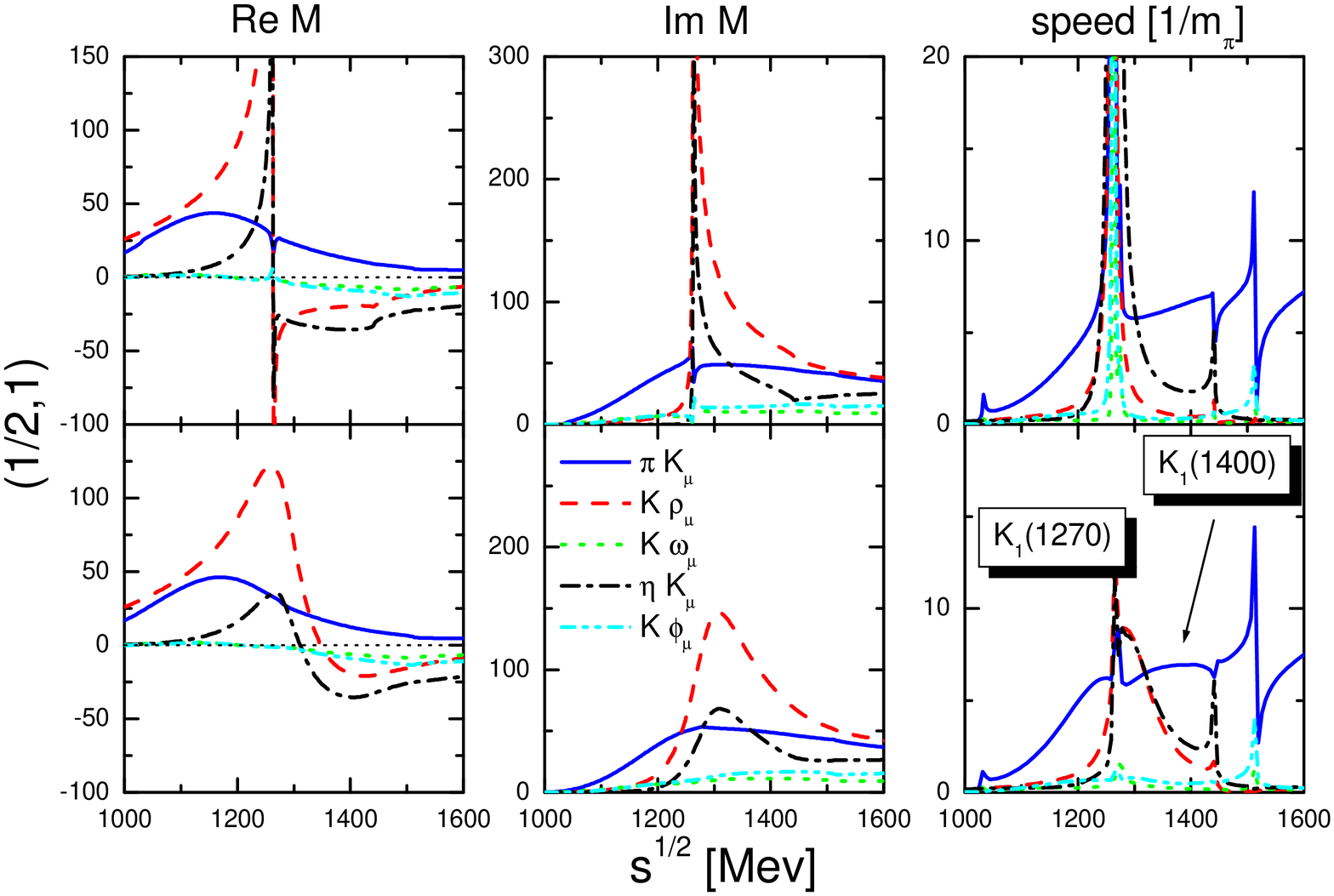}
\end{center}
\caption{Scattering amplitudes and
speeds for meson resonances with $J^P\!=\!1^+$ and $(I,S)=(\frac{1}{2},1)$ (see
(\ref{def-speed})). Parameter-free results are obtained
in terms of physical masses and $f=90$ MeV. The second row shows the effect of using
realistic spectral distributions for the $\rho_\mu$- and $K_\mu$-mesons.} \label{fig:1}
\end{figure}

\begin{figure}[t]
\begin{center}
\includegraphics[width=14.0cm,clip=true]{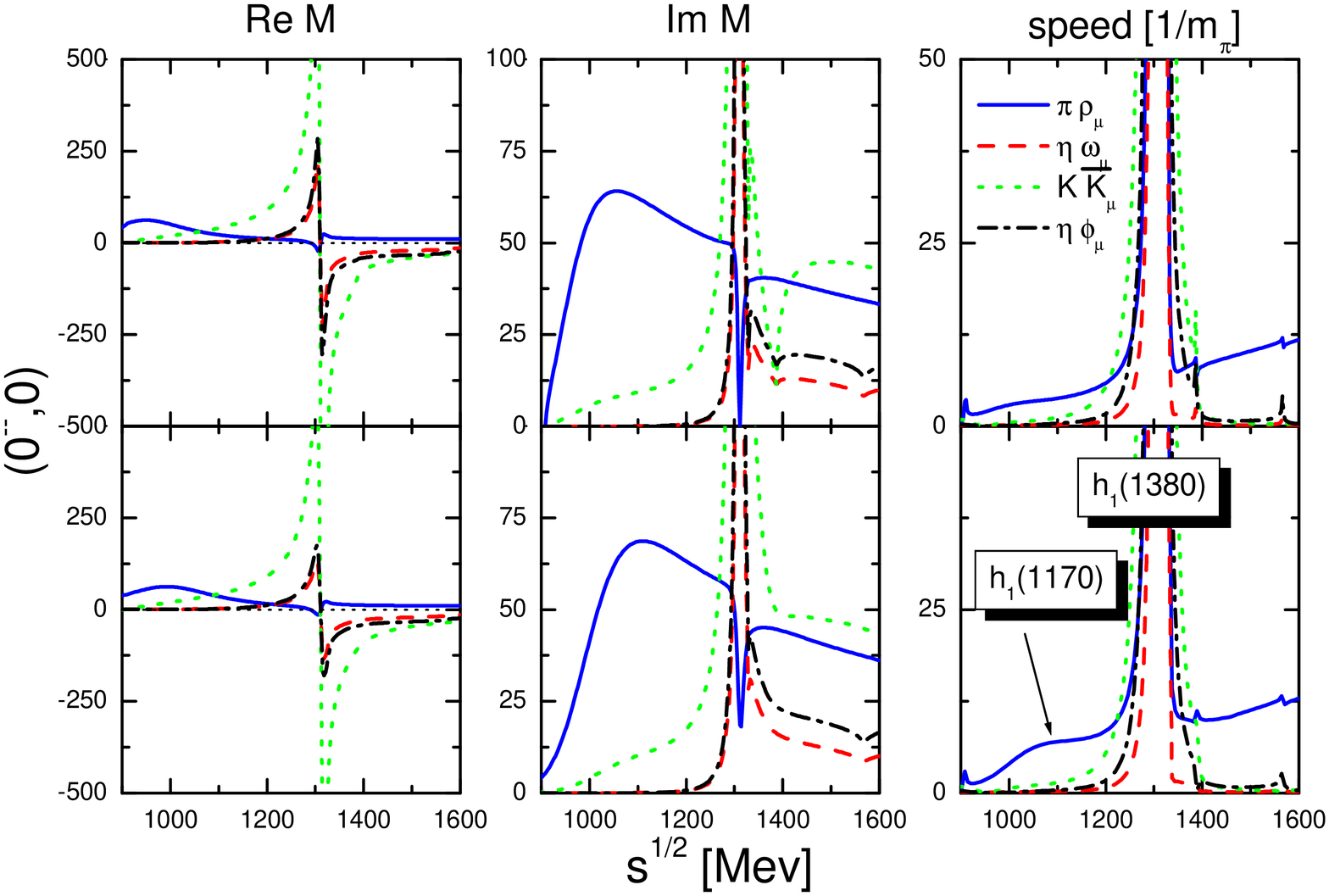}
\end{center}
\caption{Scattering amplitudes  and
speeds for meson resonances with $J^P\!=\!1^+$ and $(I^G,S)=(0^-,0)$ (see
(\ref{def-speed})). Parameter-free results are obtained
in terms of physical masses and $f=90$ MeV. The second row shows the effect of using
realistic spectral distributions for the $\rho_\mu$- and $K_\mu$-mesons.} \label{fig:2}
\end{figure}

\begin{figure}[t]
\begin{center}
\includegraphics[width=14.0cm,clip=true]{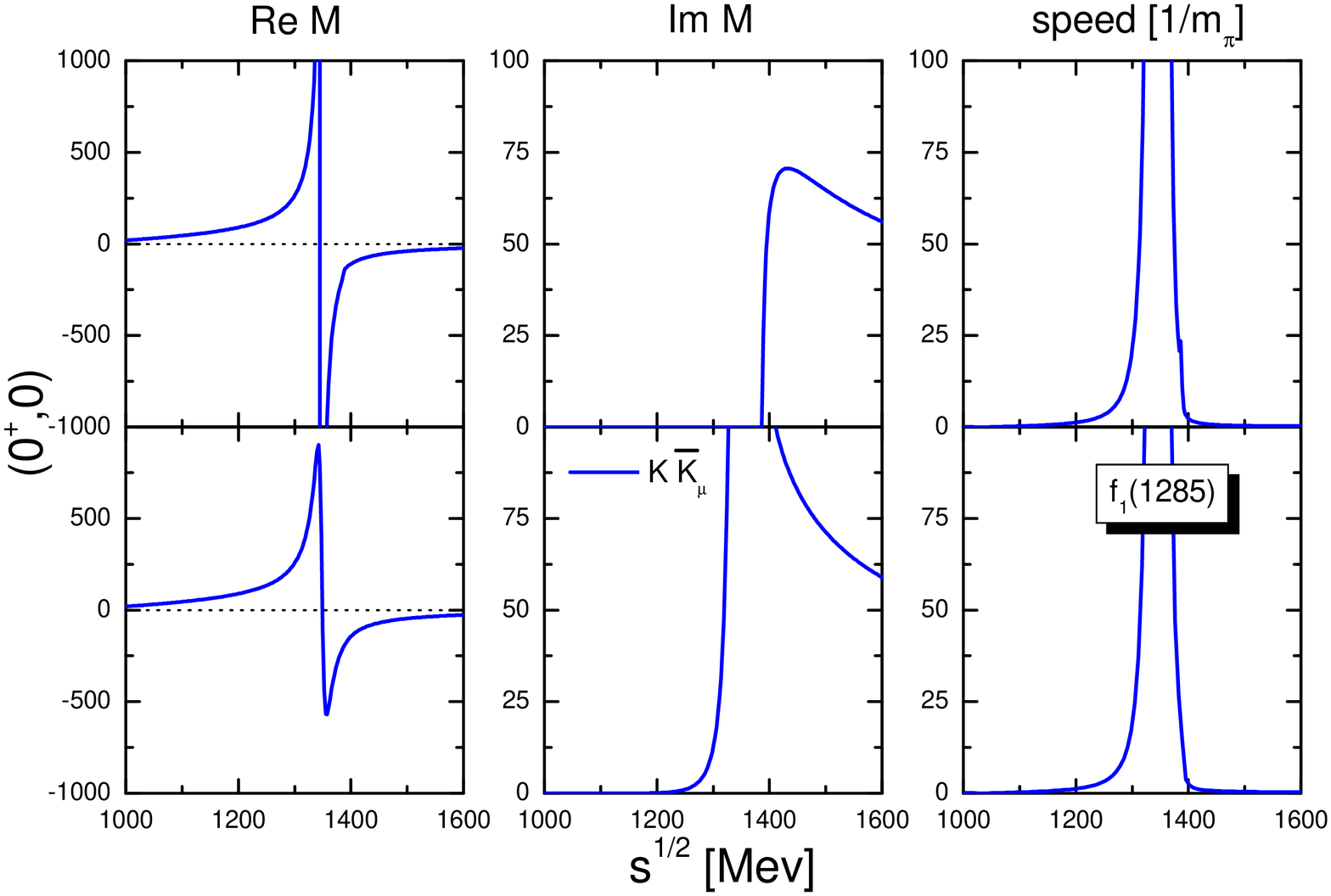}
\end{center}
\caption{Scattering amplitudes  and
speeds for meson resonances with $J^P\!=\!1^+$ and $(I^G,S)=(0^+,0)$ (see
(\ref{def-speed})). Parameter-free results are obtained
in terms of physical masses and $f=90$ MeV. The second row shows the effect of using
realistic spectral distributions for the $\rho_\mu$- and $K_\mu$-mesons.} \label{fig:3}
\end{figure}

\begin{figure}[t]
\begin{center}
\includegraphics[width=14.0cm,clip=true]{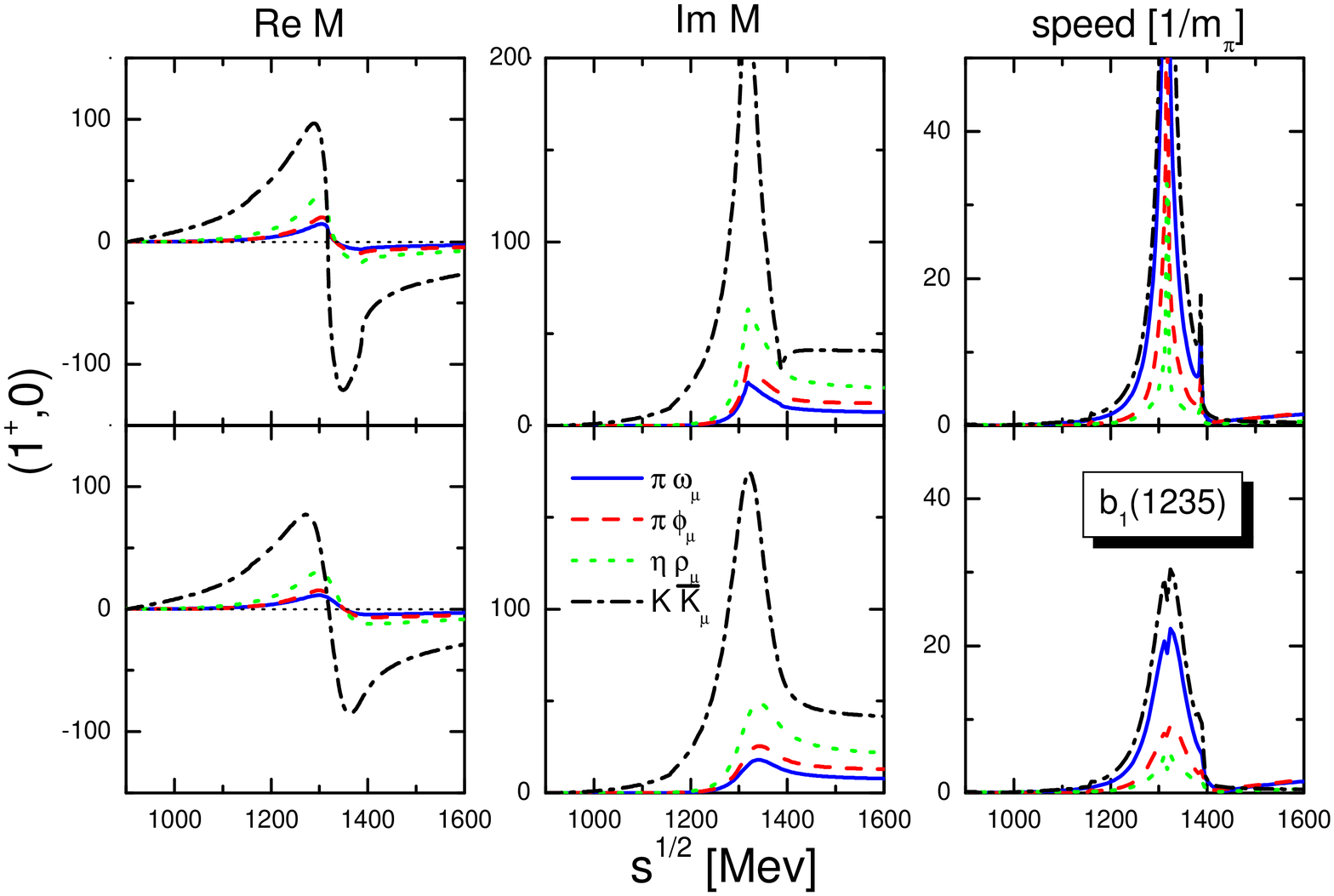}
\end{center}
\caption{Scattering amplitudes  and
speeds for meson resonances with $J^P\!=\!1^+$ and $(I^G,S)=(1^+,0)$ (see
(\ref{def-speed})). Parameter-free results are obtained
in terms of physical masses and $f=90$ MeV. The second row shows the effect of using
realistic spectral distributions for the $\rho_\mu$- and $K_\mu$-mesons.} \label{fig:4}
\end{figure}

\begin{figure}[t]
\begin{center}
\includegraphics[width=14.0cm,clip=true]{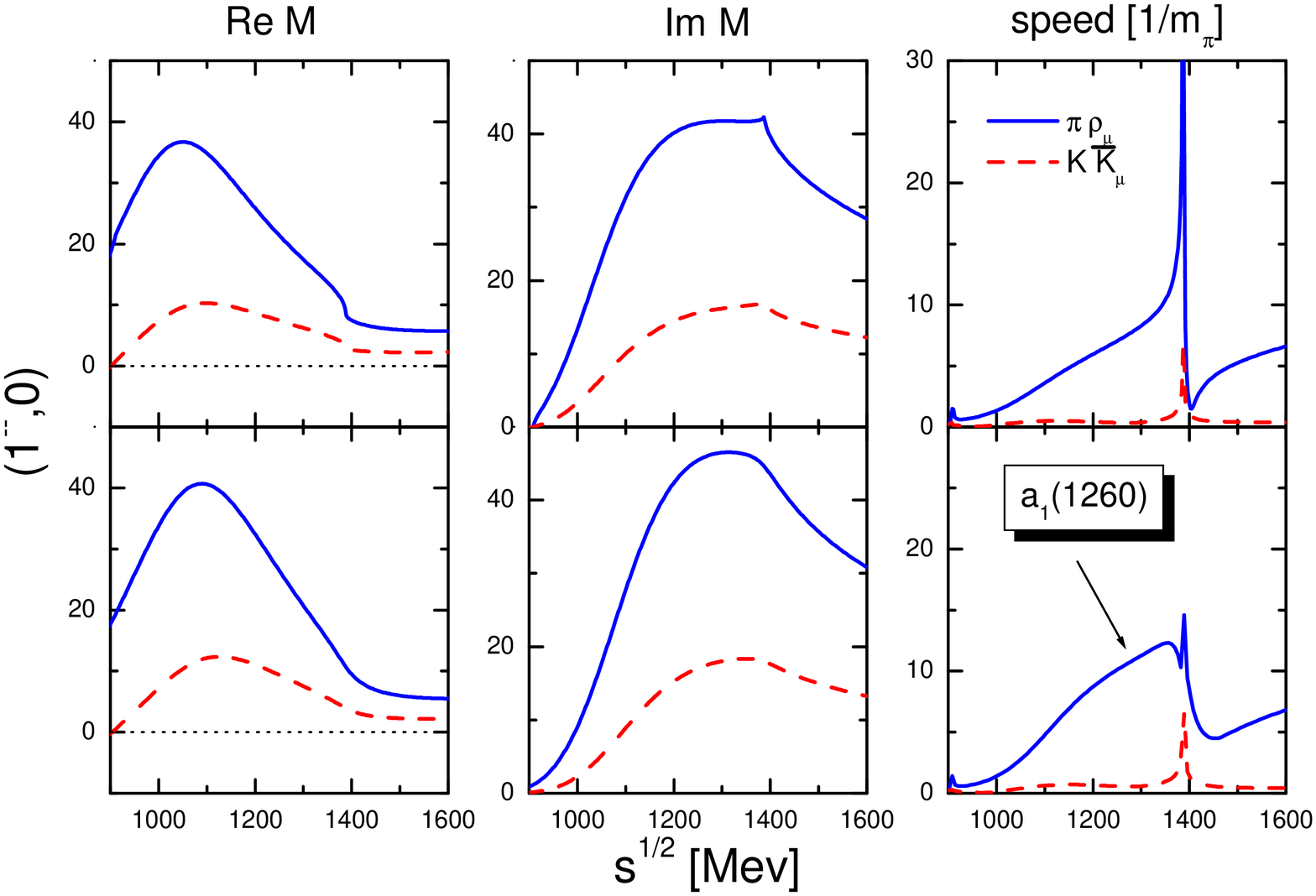}
\end{center}
\caption{Scattering amplitudes  and
speeds for meson resonances with $J^P\!=\!1^+$ and $(I^G,S)=(1^-,0)$ (see
(\ref{def-speed})). Parameter-free results are obtained
in terms of physical masses and $f=90$ MeV. The second row shows the effect of using
realistic spectral distributions for the $\rho_\mu$- and $K_\mu$-mesons.} \label{fig:5}
\end{figure}

In order to study the formation of meson resonances we generate speed plots as suggested
by H\"ohler \cite{Hoehler:speed}. The speed ${\rm Speed}_{ab}^{(I,S)}(\sqrt{s})$ of a given
channel $a\,b$, is related to the delay time \cite{speed} of a resonance produced in a
scattering experiment. It is introduced by \cite{Hoehler:speed,speed},
\begin{eqnarray}
&& t^{(I,S)}_{ab}(\sqrt{s}\,)=\frac{1}{8\,\pi \,\sqrt{s}}\,
\Big( p^{(a)}_{cm}\,N_a^{(I,S)}(\sqrt{s}\,)\,p_{cm}^{(b)}\,
N_b^{(I,S)}(\sqrt{s}\,)\Big)^{1/2}\,M_{ab}^{(I,S)}(\sqrt{s}\,)\,,
\nonumber\\
&& {\rm Speed}_{ab}^{(I,S)}(\sqrt{s}\,) = \Big|\sum_{c}\,
 \Big[\frac{d}{d \,\sqrt{s}}\, t_{ac}^{(I,S)}(\sqrt{s}\,)\Big]\,
 \Big(\delta_{cb}+2\,i\,t_{cb}^{(I,S)}(\sqrt{s}\,) \Big)^*
\Big| \,.
\label{def-speed}
\end{eqnarray}
The merit of producing speed plot lies in a convenient property of the latter allowing
a straight forward extraction of resonance parameters.
Assume that a coupled-channel amplitude $M_{ab}(\sqrt{s}\,)$ develops a pole of mass $m_R$,
with
\begin{eqnarray}
&& M_{ab}(\sqrt{s}\,) = -\frac{ g^*_a\,g^{\phantom{*}}_b\,m_R}{\sqrt{s}-m_R + i\,\Gamma/2} \,,\qquad
\Gamma_a = \frac{|g_a|^2}{4\,\pi}\,|p^{(a)}_{cm}|\,N_a(m_R)\,,
\label{def-res}
\end{eqnarray}
where the total resonance width, $\Gamma$, is given by the sum of all partial
widths. For notational simplicity we introduce the parameters,
$\Gamma_a$, also for channels that are closed, i.e. where the relative momentum
$p_{cm}^{(a)}$ in (\ref{def-res}) is imaginary. The speed plots take
a maximum  at the resonance mass $\sqrt{s}=m_R$. Some algebra
leads to the result
\begin{eqnarray}
&& {\rm Speed}_{aa}(m_R\,) =
\Bigg\{
\begin{array}{ll}
2\,\frac{\Gamma_a}{\Gamma^2}\,\Big|2\,\sum_{c}\,\frac{\Gamma_c}{\Gamma} -1\Big| \,
\qquad \qquad  \,{\rm if} \; a= {\rm open}\\
2\,\frac{\Gamma_a}{\Gamma^2}\,\Big|2\, \sum_{c}\,\frac{\Gamma_c}{\Gamma} -i\,\Big|
\, \qquad \qquad \,{\rm if} \; a= {\rm closed} \,,
\end{array}
\nonumber\\
&& \Gamma = \sum_{a={\rm open}} \Gamma_a \,,
\label{speed-an}
\end{eqnarray}
where the summation index $c$ in (\ref{speed-an}) corresponds to the one in
(\ref{def-speed}). The result (\ref{speed-an}) clearly demonstrates that the
speed \cite{speed} of a resonance in a given open channel $a$ is not only a function
of the total width parameter $\Gamma$ and the partial width $\Gamma_a$. It does depend
also on how strongly closed channels couple to that resonance. This is in contrast to the
delay time \cite{speed} of a resonance for which closed channels do not contribute.

A complementary analysis may be performed by searching for complex poles of the
scattering amplitudes on 2nd Riemann sheets. At the leading order level there is
however not much point performing such a study. The inclusion of chiral
correction terms is expected to be more important than the slightly different values
obtained for the resonances masses extracted from speed plots versus from the
position of complex poles. To guarantee good analytic properties of the scattering
amplitudes it is sufficient to check to what extent the scattering amplitudes
satisfy subtracted dispersion-integral representations. An unphysical singularity within
the applicability domain would invalidate such a representation. The absence of unphysical
structures is also excluded to a large extent by the form of real and imaginary parts of
the amplitudes. The resonance like behaviour of all amplitudes is a strong indication that
there are no spurious singularities that are relevant, i.e. in the applicability domain.

To explore the multiplet structure of the resonance states we  study first the
'heavy' SU(3) limit \cite{Granada,Copenhagen} with
$m_{\pi, K, \eta} \simeq 500$ MeV and $m_{\rho, \omega, K^*, \phi}\simeq 900$
MeV. In this case all resonance states turn into bound states forming two degenerate octets and
one singlet of the SU(3) flavor group with masses  1367 MeV and 1289 MeV respectively.
The latter numbers are quite insensitive to the precise value of the subtraction scale.
For instance increasing (decreasing) the subtraction scale by 20 $\%$ away from its natural
value the octet bound-state mass comes at 1383 MeV (1353 MeV). Our result is a direct
reflection of the Weinberg-Tomozawa interaction,
\begin{eqnarray}
8 \otimes 8 = 1 \oplus 8 \oplus 8 \oplus 10 \oplus \overline{10}\oplus 27 \,,
\label{}
\end{eqnarray}
which predicts attraction in the two octet and the singlet channel. This finding
is analogous to the results of \cite{Granada,Jido03} that found two degenerate octet
and one singlet state in the SU(3) limit of meson-baryon scattering with $J^P\!=\!\frac{1}{2}^-$.
Taking the 'light' SU(3) limit \cite{Copenhagen} with $m_{\pi, K, \eta} \simeq 140$ MeV
and $m_{\rho, \omega, K^*, \phi}\simeq 700$ MeV we do not observe any
bound-state nor resonance signals anymore. A further interesting limit to study
is $N_c \to \infty$. Since the scattering kernel is proportional to the $f^{-2}\sim N_c^{-1}$
the interaction strength vanishes in that limit and no resonances are generated.

In Fig. \ref{fig:0} we demonstrate the quality of the proposed matching procedure as
applied for the forward scattering amplitudes describing the $\pi \,K_\mu$ scattering
process in the 'heavy' SU(3) limit. It is shown the term in front of the $g_{\mu \nu}$
structure of the scattering amplitude $T_{\mu \nu}$ as a function of $\sqrt{s}$, where all
but the s-wave contributions are evaluated at tree-level for simplicity. The figure
clearly illustrates the smooth matching of s-channel and u-channel iterated amplitudes at
subthreshold energies. Modifying the subtraction scale by about $10\%$ in either direction
does not deteriorate the quality of the matching. It should be pointed out that once chiral
correction terms are included in the calculation the quality of the matching is expected to
further improve.

Figs. \ref{fig:1}-\ref{fig:5} show
the resonance spectrum that arises using physical masses (first row) and using
realistic spectral distributions for the broad vector mesons (second row).
Clear signals in the speed plots of the $({\textstyle{1\over 2}},\pm 1)$,
$(0^\pm,0)$ and $(1^\pm,0)$ channels are seen. No resonance is
found in the remaining channels. The resonances can be unambiguously identified with
the axial-vector meson resonances ($h_1(1170), h_1(1380), f_1(1285), a_1(1260), b_1(1235),
K_1(1270), K_1(1400)$).

In the 'heavy' SU(3) limit the $({\textstyle{1\over 2}},\pm 1)$ channel
shows two bound states reflecting the presence of two degenerate octet states.
Using physical masses the degeneracy is lifted as illustrated in Fig. \ref{fig:1}
and a narrow state at 1263 MeV and a broad state at about 1300 MeV arise.  The resonance
masses are determined from the maxima of the speed where one has to discard the narrow structures
induced by the square root singularities at the various thresholds. Thus, if a resonance is close
to a threshold it is difficult to read off its properties from the speed plots. The simple results
(\ref{def-res},\ref{speed-an}) can not be applied. The effect of using
realistic spectral distributions for the intermediate $\rho_\mu$- and $K_\mu$-mesons
is demonstrated in the second row of Fig. \ref{fig:1}, the first row showing results with
sharp vector meson masses. The resonance signal in the speed plots becomes much clearer
since using spectral distributions for the broad intermediate states smears away the
square-root singularities in the speeds at the corresponding thresholds.
In this case we introduce the speed (\ref{def-speed})
with respect to the invariant amplitudes $M(\sqrt{s}\,)$ evaluated in terms of
loop functions folded with spectral distributions of the intermediate states, but
use $p_{cm}^{(a)}$ and $p_{cm}^{(b)}$ in (\ref{def-speed}) defined with respect to
sharp masses. Thus the parameters $\Gamma_a$ introduced in (\ref{def-res})
characterizing the maximum of the speed (\ref{speed-an}) determine the coupling constants
$g_a$ via (\ref{def-res}) and not the partial-decay width in a channel $a$ where broad
intermediate states are used. In contrast, the parameter $\Gamma$ has the interpretation of the
total width, i.e. $\Gamma \neq \sum_{a={\rm open}} \Gamma_a$ in this case.
Our result is quite consistent with the empirical properties of the $K_1(1270)$ meson. It has
a width of about 90 MeV and decays dominantly into the $K\,\rho_\mu-$channel \cite{PDG02}.
The second much broader state is assigned a width of about 175 MeV
resulting almost exclusively from its decay into the $\pi \,K_\mu-$channel \cite{PDG02}.

Similarly, in the heavy SU(3) limit the $(0^-,0)$ channel
shows two bound states associated with a singlet and an octet state. Using physical
masses a broad state at about 1100 MeV and a narrow state at 1303 MeV should be
identified with the $h_1(1170)$ and $h_1(1380)$ resonance (see Fig. \ref{fig:2}).
Here we assign the
$h_1(1380)$-resonance, for which its quantum numbers except its parity and angular momentum
$J^P=1^+$ are unknown, the isospin and G-parity quantum numbers $I^G=0^-$.
This is a clear prediction of the chiral coupled-channel dynamics.
The latter state has so far been seen only through its decay into the
$K\,\bar K_\mu$- and $\bar K\,K_\mu -$channels \cite{PDG00}.
Its small width of about 80 MeV \cite{PDG00} is consistent with the narrow structure
seen in Fig. \ref{fig:2}. The second resonances state in Fig. \ref{fig:2} is most clearly
seen in the $\pi \rho_\mu-$channel. This is consistent with the empirical properties of
the $h_1(1170)$ resonance which so far has been seen only through its $\pi \rho_\mu-$decay
leading to a large width of about 360 MeV \cite{PDG02}.

The $(0^+,0)$-speed (see Fig. \ref{fig:3}) shows a bound-state at
mass 1341 MeV a value somewhat above the mass of the $f_1(1285)$ resonance.
Using a spectral distribution for the $K_\mu(892)$ in the
intermediate states $K\,\bar K_\mu$ and $\bar K\,K_\mu $ states
a narrow resonance appears. Its width of about 10 MeV is a factor two smaller
than the empirical value \cite{PDG02}.
The $(1^+,0)$-speed of Fig. \ref{fig:4} shows a resonance at 1310 MeV
to be identified with the $b_1(1235)$ resonance. From the maximum of the imaginary
part of the scattering amplitudes at the resonance peak one can directly read off ratios of
coupling constants. Fig. \ref{fig:4} clearly demonstrates that the smallest coupling
constant is predicted for the $\pi\, \omega_\mu -$channel. Nevertheless, the hadronic
decay of the $b_1(1235)$ is completely dominated by the $\pi\,\omega_\mu-$channel. This
is a simple consequence of phase-space kinematics. The widths of the resonance as
indicated by Fig. \ref{fig:4} is quite compatible with the empirical value of
about 140 MeV \cite{PDG02}.  The $a_1(1260)$ resonance is found
in the $(1^-,0)$-speed of Fig. \ref{fig:2} as a broad peak with a mass of about 1300 MeV.
Empirically its width is estimated to be about 250-600 MeV \cite{PDG02} resulting from its
decay into the $\pi\,\rho-$channel.

The structure of the $h_1(1380),f_1(1285)$ and $b_1(1235)$ resonances as predicted by chiral
coupled-channel dynamics is quite intriguing since those resonances couple dominantly to the
$K\,\bar K_\mu-$channel. This implies
that the latter channel is the driving force that generates these resonances dynamically.
This finding is very much analogous to the structure of the scalar $f_0(980)$ resonance that
strongly couples to the $K \bar K-$channel and emphasizes the importance of the chiral SU(3)
symmetry even for non-strange resonances. It should be emphasized that the results obtained
here at leading order can be improved further by incorporating chiral correction terms into
the analysis. In view of the remarkable success of the leading order Weinberg-Tomozawa
interaction one would expect a rapidly converging expansion.

We conclude with an interesting by-product of our analysis. The s-wave scattering lengths of pions
with vector mesons are predicted. The scattering length of  a pseudo-scalar meson (P)
of mass $m$ off a vector meson (V) off mass $M$ is identified,
\begin{eqnarray}
4\,\pi \Big(1+ \frac{m}{M}\Big)\,a^{(I,S)}_{P V}
= \frac{3}{4\,M}\, M^{(I,S)}(m+M) \,.
\label{def-asw}
\end{eqnarray}
At leading order with $M^{(I,S)}(\sqrt{s}\,) \to V^{(I,S)}(\sqrt{s}\,) $ we recover the
Weinberg-Tomozawa theorem,
\begin{eqnarray}
&& 4\,\pi \Big(1+ \frac{m}{M}\Big)\,a^{(I,S)}_{P V} = \frac{m}{2\,f^2}\,C^{(I,S)}
+ {\mathcal O}\left( m^2 \right) \,,
\nonumber\\
&& a_{\pi \,\rho_\mu} =
\frac{1}{9}\,\Big(a^{(0,0)}_{\pi \rho_\mu }+3\,a^{(1,0)}_{\pi \,\rho_\mu }
+5\,a^{(2,0)}_{\pi \,\rho_\mu } \Big) = 0\,,
\qquad a_{\pi \,\omega_\mu } =
\frac{1}{3}\,a^{(1^+,0)}_{\pi \,\omega_\mu } =0 \,,
\nonumber\\
&& a_{\pi \,\phi_\mu } =
\frac{1}{3}\,a^{(1^+,0)}_{\pi \,\phi_\mu} =0 \,,\qquad
a_{\pi K_\mu} =
\frac{1}{3}\,\Big(a^{(\frac{1}{2},\pm 1)}_{\pi\, K_\mu}
+2\,a^{(\frac{3}{2},\pm 1)}_{\pi\, K_\mu}\Big) =0\,,
\label{WT-theorem}
\end{eqnarray}
\tabcolsep=1.3mm
\renewcommand{\arraystretch}{1.5}
\begin{table}[t]\begin{center}
\begin{tabular}{|c|c|c|c|c|c||c|c|c|c|}
\hline
[fm] & $a^{(0,0)}_{\pi \rho_\mu}$ & $a^{(1,0)}_{\pi \,\rho_\mu} $ & $a^{(2,0)}_{\pi \,\rho_\mu} $
& $a^{(\frac{1}{2}, \pm 1)}_{\pi K_\mu} $ & $a^{(\frac{3}{2}, \pm 1)}_{\pi K_\mu} $ &
$a_{\pi \rho_\mu} $ & $a_{\pi \omega_\mu} $ & $a_{\pi \phi_\mu} $ & $a_{\pi K_\mu} $
\\
\hline \hline
WT  & 0.45 &  0.23  & -0.23  & 0.23  &  -0.12  &  0  & 0   & 0 & 0
 \\  \hline

$\chi-BS(3)$  & 0.69 &  0.27 & -0.20 & 0.29 & -0.10 & 0.06 & 0.00  & 0.01 & 0.09
 \\  \hline
\end{tabular}
\end{center}
\caption{S-wave scattering length $a_{\pi V}$ with $V=( \rho_\mu, \omega_\mu, \phi_\mu, K_\mu)$.
The first row gives the leading order prediction of Weinberg and Tomozawa
(see (\ref{WT-theorem})). These results are confronted in the second row with the scattering length as evaluated
in the $\chi-BS(3)$-scheme. Here we use sharp vector meson masses.}
\label{scattering-length:tab}
\end{table}
which predicts that all isospin averaged scattering lengths of a pion off any vector meson
vanish at leading order. In Tab. \ref{scattering-length:tab} the scattering lengths as
predicted by Weinberg and Tomozawa are confronted with the values obtained form the chiral
coupled-channel theory. The deviations obtained are significant in the
$(0,0)$ and $(\frac{1}{2},\pm 1)$ channels leading to non-zero and attractive
isospin averaged scattering length for the $\rho_\mu-$ and $K_\mu-$mesons. In the
table we present the scattering length obtained for sharp vector-meson masses.
In the more realistic case of broad states the scattering lengths are not defined anymore
unambiguously. If we use (\ref{def-asw}) as a definition with sharp values for the mass
parameter $M$ but using the amplitudes $M^{(I,S)}(\sqrt{s})$ evaluated in terms of broad
intermediate states, the numbers in Tab. \ref{scattering-length:tab} change somewhat.
In particular the isospin averaged $\pi \rho-$scattering length is reduced to 0.02 fm.

{\bfseries{Acknowledgments}}

M.F.M.L. acknowledges stimulating discussions with M.A. Nowak.

\newpage

\section{Appendix}

In this appendix we construct the projectors, ${\mathcal Y}^{(J P)}_{ab}(\bar q,q;w)$,
introduced in (\ref{t-exp}). The latter define the important notion of 'on-shell'
irreducibility. Consider the on-shell scattering amplitude of a pseudo-scalar
meson $P(q)$ off a vector meson $V(p,\lambda)$ with polarization $\lambda$,
\begin{eqnarray}
&&\langle P(\bar q)\,V(\bar p, \bar \lambda)| \,T\,| P(q)\,V(p, \lambda) \rangle
\nonumber\\
&& \qquad = (2\pi)^4
\, \delta^4(q+p-\bar q-\bar p)\,\epsilon^\dagger_\mu(\bar p,\bar \lambda)\,
T^{\mu \nu}(\bar q, q;w)\,\epsilon_\nu(p,\lambda)\,,
\label{}
\end{eqnarray}
where we suppress isospin and strangeness quantum numbers for simplicity.
The scattering amplitude is subject to various constraints. Covariance
together with parity and time reversal conservation lead to a representation of
$T_{\mu \nu}(\bar q,q;w)$ in terms of five scalar amplitudes $F_{i=1,..,5}$, and a complete set of
Lorentz tensors $L_i^{\mu \nu}$,
\begin{eqnarray}
&& T^{\mu \nu}(\bar q,q;w) = \sum_{i=1}^5\,F_i(\bar q,q;w)\,L^{\mu \nu}_i(\bar q,q;w)
\\
&& L_1^{\mu \nu} = g^{\mu \nu}- \frac{w^\mu\,w^\nu}{w^2}\,, \quad
L_2^{\mu \nu} =\frac{w^\mu\,w^\nu}{w^2}\,,\quad
L_3^{\mu \nu} =  \frac{w^\mu}{\sqrt{w^2}}\,\Big(\bar q^\nu - \frac{\bar q\cdot w}{w^2}\,w^\nu \Big)\,, \qquad
\nonumber\\
&& L_4^{\mu \nu} =\Big(q^\mu- \frac{q\cdot w}{w^2}\,w^\mu \Big)\,\frac{w^\nu}{\sqrt{w^2}}\,,\!\!\quad
L^{\mu \nu}_5= \Big(q^\mu- \frac{q\cdot w}{w^2}\,w^\nu \Big)\,
\Big(\bar q^\nu - \frac{\bar q\cdot w}{w^2}\,w^\nu \Big)\,. \nonumber
\label{t-decomp}
\end{eqnarray}
A further important constraint follows from the two-particle unitarity condition which is
efficiently implemented in terms of helicity states \cite{Jacob:Wick}. In the center of
mass frame with $w_\mu = (\sqrt{s},\vec 0)$, helicity matrix elements of
the scattering amplitude are decomposed into
partial wave amplitudes, $\langle \bar \lambda |T^{(J)}| \lambda \rangle$,
of given total angular momentum $J$,
\begin{eqnarray}
&& \epsilon^\dagger_\mu(\bar p, \bar \lambda)\,T^{\mu \nu} (\bar q, q;w)\,\epsilon_\nu(p,\lambda)
= \sum_J\,(2\,J+1)\,\langle \bar \lambda |T^{(J)}| \lambda \rangle \,d^{(J)}_{\lambda \bar \lambda}(\theta )
\nonumber\\
&& \epsilon_\mu (p) =
\left(
\begin{array}{c}
0 \\
\frac{\mp 1}{\sqrt{2}}\\
\frac{-i}{\sqrt{2}}\\
0
\end{array}
\right)\,,
\left(
\begin{array}{c}
\frac{p_{\rm cm}}{M} \\
0\\
0\\
\frac{\omega}{M}
\end{array}
\right)\,,\quad
\epsilon_\mu (\bar p) =
\left(
\begin{array}{c}
0 \\
\frac{\mp \cos \theta}{\sqrt{2}} \\
\frac{-i}{\sqrt{2}}\\
\frac{\pm \sin \theta}{\sqrt{2}}
\end{array}
\right)\,,
\left(
\begin{array}{c}
\frac{p_{\rm cm}}{M} \\
\frac{\omega}{M} \,\sin \theta \\
0\\
\frac{\omega}{M}\,\cos \theta
\end{array}
\right)\,,
\label{helicity-exp}
\end{eqnarray}
with
$\epsilon_\mu(p)=\epsilon_\mu(p,\pm 1),\epsilon_\mu(p,0)$,
$\epsilon_\mu(\bar p)=\epsilon_\mu(\bar p,\pm 1),\epsilon_\mu(\bar p,0)$ and
the scattering angle $\theta  $. The objects $d^{(J)}_{\lambda \lambda}(\theta)$ are
Wigner's rotation functions and $\omega = (M^2+p_{\rm cm}^2)^{1/2}$.
The unitarity constraint now takes the simple form
\begin{eqnarray}
t_{\bar \lambda \lambda}^{(J)} =\langle \bar \lambda |T^{(J)}| \lambda \rangle \,,
\qquad
\Im( t^{(J)})^{-1}_{\bar \lambda \lambda} =
\frac{p_{\rm cm}}{8\pi\,\sqrt{s}}\,\delta_{\bar \lambda \lambda} \,.
\label{}
\end{eqnarray}

For the case at hand the channels $J^{P=+}$ and
$J^{P=-}$ lead to two-dimensional and one-dimensional projectors respectively. In principal
the form of the projectors follows from a boost of the representation (\ref{helicity-exp}).
However, it is not straight forward to boost partial wave amplitudes. In general this
task can be quite tedious. Naive prescriptions
typically lead to kinematical singularities and must therefore be rejected. The
precise form of the projectors will be derived in the following. In a first step
the invariant amplitudes $F_i$ are expressed in term of helicity matrix elements, $\phi_i$,
of the scattering amplitude,
\begin{eqnarray}
&& \phi_1 = \epsilon^\dagger_\mu(\bar p,+1)\,
T^{\mu \nu}(\bar q, q;w)\,\epsilon_\nu(p,+1) +\epsilon^\dagger_\mu(\bar p,+1)\,
T^{\mu \nu}(\bar q, q;w)\,\epsilon_\nu(p,-1) \,,
\nonumber\\
&& \phi_2 = \epsilon^\dagger_\mu(\bar p,+1)\,
T^{\mu \nu}(\bar q, q;w)\,\epsilon_\nu(p,+1) -\epsilon^\dagger_\mu(\bar p,+1)\,
T^{\mu \nu}(\bar q, q;w)\,\epsilon_\nu(p,-1) \,,
\nonumber\\
&& \phi_3 = \frac{1}{\sin \theta}\,\epsilon^\dagger_\mu(\bar p,0)\,
T^{\mu \nu}(\bar q, q;w)\,\epsilon_\nu(p,+1)\,,
\nonumber\\
&& \phi_4 = \frac{1}{\sin \theta}\,\epsilon^\dagger_\mu(\bar p,+1)\,
T^{\mu \nu}(\bar q, q;w)\,\epsilon_\nu(p,0)\,,
\nonumber\\
&& \phi_5 = \epsilon^\dagger_\mu(\bar p,0)\,
T^{\mu \nu}(\bar q, q;w)\,\epsilon_\nu(p,0) \,.
\label{def-phi}
\end{eqnarray}
The five invariant amplitudes can be expressed in terms of the helicity amplitudes
$\phi_i$,
\begin{eqnarray}
\left(
\begin{array}{c}
F_1 \\
F_2\\
F_3 \\
F_4 \\
F_5
\end{array}
\right)
=
\left(
\begin{array}{ccccc}
-1 & 0 & 0 & 0& 0\\
\frac{\omega\,\bar \omega\,x}{p_{\rm cm}\,\bar p_{\rm cm}\,(1-x^2)} & -\frac{\omega\,\bar \omega\,x^2}{p_{\rm cm}\,\bar p_{\rm cm}\,(1-x^2)} & \frac{\sqrt{2}\,\omega\,\bar M\,x}{p_{\rm cm}\,\bar p_{\rm cm}}  & -\frac{\sqrt{2}\,\bar \omega\,M\,x}{p_{\rm cm}\,\bar p_{\rm cm}}& \frac{M\,\bar M}{p_{\rm cm}\,\bar p_{\rm cm}}\\
\frac{\bar \omega}{\bar p_{\rm cm}^2\,(1-x^2)} & -\frac{\bar \omega\,x}{\bar p_{\rm cm}^2\,(1-x^2)} & \frac{\sqrt{2}\,\bar M}{\bar p_{\rm cm}^2} & 0& 0\\
\frac{\omega}{p_{\rm cm}^2\,(1-x^2)} & -\frac{\omega\,x}{p_{\rm cm}^2\,(1-x^2)} & 0 & -\frac{\sqrt{2}\,M}{p_{\rm cm}^2} & 0\\
\frac{x}{p_{\rm cm}\,\bar p_{\rm cm}\,(1-x^2)}\, &  \frac{-1}{p_{\rm cm}\,\bar p_{\rm cm}\,(1-x^2)} &0 & 0 & 0
\end{array}
\right)
\left(
\begin{array}{c}
\phi_1 \\
\phi_2 \\
\phi_3 \\
\phi_4 \\
\phi_5
\end{array}
\right)\,,\nonumber\\
\label{res-f}
\end{eqnarray}
where we discriminated the masses and relative momenta of the initial and final states with
$M, \bar M$ and $p_{\rm cm}, p_{\rm cm}'$. Furthermore we use  $x = \cos \theta$ and
$\omega = (M^2+p_{\rm cm}^2)^{1/2}$ and $\bar \omega = (\bar M^2+\bar p_{\rm cm}^2)^{1/2}$.
According to the general decomposition (\ref{helicity-exp}) the amplitudes $\phi_i$ can
be expressed in terms of partial wave helicity amplitudes
$\langle \bar \lambda | T^{(J)} |\lambda  \rangle $,
\begin{eqnarray}
&& \phi_1 =
\sum_J \,\frac{2\,J+1}{J\,(J+1)}\,
\Bigg( \langle 1_+ | T^{(J)} | 1_+\rangle \,P_J'(\cos \theta)
\nonumber\\
&& \qquad \qquad \qquad \qquad - \langle 1_- | T^{(J)} | 1_-\rangle
\big(\cos \theta \,P_J'(\cos \theta) - J\,(J+1)\, P_J(\cos \theta) \Big)
\Bigg)\,,
\nonumber\\
&& \phi_2=
\sum_J \,
\,\frac{2\,J+1}{J\,(J+1)}\,
\Bigg( \langle 1_- | T^{(J)} | 1_-\rangle \,P_J'(\cos \theta)
\nonumber\\
&& \qquad \qquad \qquad \qquad -\langle 1_+ | T^{(J)} | 1_+\rangle \,\big(\cos \theta \,P_J'(\cos \theta) - J\,(J+1)\, P_J(\cos \theta) \Big)
\Bigg)\,,
\nonumber\\
&& \phi_3 =
-\frac{1}{\sqrt{2}}\,\sum_J \,\frac{2\,J+1}{\sqrt{J\,(J+1)}}\,\, \langle 0 | T^{(J)} | 1_+\rangle \,
P'_J(\cos \theta) \,,
\nonumber\\
&& \phi_4 =
+\frac{1}{\sqrt{2}}\,\sum_J \,\frac{2\,J+1}{\sqrt{J\,(J+1)}}\, \langle 1_+ | T^{(J)} | 0 \rangle \,P'_J(\cos \theta) \,,
\nonumber\\
&& \phi_5 =
+\sum_J \,(2\,J+1)\, \langle 0 | T^{(J)} | 0 \rangle \,P_J(\cos \theta) \,,
\label{res-phi}
\end{eqnarray}
where we introduced parity eigenstates with
$\langle 1_\pm ,J| = (\pm \langle +1,J | + \langle -1,J |)/\sqrt{2}$ and
applied the useful identities \cite{Var}
\begin{eqnarray}
&& d^{(J)}_{00} (\theta) = P_J(\cos \theta) \,, \qquad
d^{(J)}_{\pm 10}(\theta) = \mp \frac{\sin \theta }{\sqrt{J\,(J+1)}}\,P_J'(\cos \theta)=-
d^{(J)}_{0\pm 1}(\theta)\,,
\nonumber\\
&& d^{(J)}_{\pm 11}(\theta) = \frac{1\pm \cos \theta}{J\,(J+1)}\,
\Big( P_J'(\cos \theta) \mp (1 \mp \cos \theta)\, P_J''(\cos \theta)\Big)=
d^{(J)}_{1\pm 1}(\theta) \,.
\label{}
\end{eqnarray}
It now appears straightforward to construct the projectors,
${\mathcal Y}^{(J P_-)}(\bar q,q;w)$, associated
with $\langle 1_- |$ and parity $P_-=(-1)^J$. It is a
single-channel projector. However, it is important to
properly boost the results (\ref{t-decomp},\ref{res-f},\ref{res-phi}) obtained in
the center of mass frame. As was pointed out in \cite{LK02} it is incorrect to
identify always
\begin{eqnarray}
&& \cos \theta  \leftrightarrow \frac{Y_{q\bar q}}{\sqrt{Y_{qq}\,Y_{\bar q \bar q}}} \,,
\label{wrong-def}
\end{eqnarray}
with
\begin{eqnarray}
&& Y_{q q} = \frac{(w\cdot q)^2}{w^2}- q^2\,, \qquad
Y_{\bar q \bar q} = \frac{(w\cdot \bar q)^2}{w^2}- \bar q^2\,,\qquad
\nonumber\\
&& Y_{q \bar q} = \frac{(w\cdot q)\,(w \cdot \bar q)}{w^2}- q\cdot \bar q\,.
\label{def-y}
\end{eqnarray}
It is clear, that if (\ref{wrong-def}) is used for instance in $P_J(\cos \theta)$, kinematical
singularities at $Y_{qq} =0 $ or $Y_{\bar q \bar q} =0 $ arise that are unphysical. The latter
are realized at the off-shell surfaces defined by $(p\cdot q)^2 =p^2\,q^2$ and
$(\bar p\cdot \bar q)^2 =\bar p^2\,\bar q^2$. Therefore the naive prescription (\ref{wrong-def})
must be rejected. It  would spoil the analytic properties of the scattering amplitude.
In order to proceed it is necessary to boost objects only that posses a proper frame independent
representation. For instance this is the case for $(p_{\rm cm}\,p_{\rm cm}')^n\,P_n(\cos \theta)$.
We identify
\begin{eqnarray}
&& (p_{\rm cm}\,p_{\rm cm}')^n\,P_n(\cos \theta) \to
\sum_{k=0}^n\,c^{(n)}_k\,
\,Y^{(n-k)/2}_{qq}\,Y^{(n-k)/2}_{\bar q\bar q}\,Y^k_{q\bar q} \,,
\nonumber\\
&& P_n(x)= \sum_{k=0}^n \,c^{(n)}_k\,x^k\,,\qquad c^{(n)}_{n-2\,r} = (-1)^r\,
\frac{(2\,n-2\,r) \,!}{2^n\,r\,!\,(n-r)\,! \,(n-2\,r)\,!}  \,.
\label{cor-id}
\end{eqnarray}
Note that the combination $n-k$ in (\ref{cor-id}) is always even and positive.
Thus no square-root singularities are picked up in (\ref{cor-id}).
An analogous replacement is applicable for $(p_{\rm cm}\,p_{\rm cm}')^n\,x\,P'_n(x)$.
We derive
\begin{eqnarray}
&& {\mathcal Y}^{(J P_-)}_{\mu \nu}(\bar q,q;w) = \sum_{k=1}^{J}
\frac{2\,J+1}{J\,(J+1)}\,c^{(J)}_{k}\,k\,(k-1)\,
 Y^{(J-k)/2}_{qq}\, Y^{(J-k)/2}_{\bar q\bar q} \,Y^{k-1}_{q\bar q}
\nonumber\\
&& \quad \times \Bigg( \Big(\frac{1}{k-1}-\frac{J\,(J+1)}{k\,(k-1)} \Big)\,
Y_{q \bar q}\,L_{1,\mu \nu}
- \frac{(\bar q \cdot w)\,(w \cdot q)}{w^2}\,L_{2,\mu \nu}
\nonumber\\
&& \qquad - \frac{w \cdot\bar  q}{\sqrt{w^2}}\,\frac{Y_{q q}}{Y_{q \bar q}}\,\,L_{3, \mu \nu}
- \frac{w \cdot q}{\sqrt{w^2}}\,\frac{Y_{\bar q \bar q}}{Y_{q \bar q}}\,L_{4, \mu \nu}
  - \frac{k}{k-1}\,L_{5,\mu \nu}
\Bigg) \,,
\nonumber\\
&& P_- = (-1)^J \,,
\label{y-minus}
\end{eqnarray}
where the coefficients $c_k^{(J)}$ are given in (\ref{cor-id}) and the
Lorentz tensors $L_{i,\mu \nu}$ were introduced in (\ref{t-decomp}).
The projector is introduced with respect to
\begin{eqnarray}
&& | 1^{(-)}_c, J \rangle = p_{\rm cm}^{J}\,| 1_- ,J\rangle \,,
\label{non-uni-def:a}
\end{eqnarray}
rather than $| 1_- ,J\rangle $.  This rescaling provides the necessary phase space factor
$(p_{\rm cm}\,\bar p_{\rm cm})^J$ required for the proper definition of the projector.
We emphasize that
the projectors ${\mathcal Y}^{(J P_-)}_{\mu \nu}(\bar q,q;w)$ are regular at the kinematical
surfaces defined by $(p\cdot q)^2 =p^2\,q^2$ and
$(\bar p\cdot \bar q)^2 =\bar p^2\,\bar q^2$. The singularity at $w^2=0$ can be avoided
by rescaling the projectors by an appropriate power in $\sqrt{w^2}$. Since the kinematical
point $w^2 =0$ is far outside the region where we will be using the projectors this is not
an issue here.

We continue and derive the projectors for the remaining sector. Here an additional
complication arises. If one tries to introduce a projector matrix with respect to the
helicity states $(\langle 1_+ ,J|, \langle 0,J|)$ it is impossible to arrive at a result that
is free of kinematical singularities. A non-unitary transformation is required to new states
$(\langle 1^{(+)}_c ,J|, \langle 2^{(+)}_c,J|)$ with
respect to which projectors can be obtained that are free of kinematical singularities,
\begin{eqnarray}
&& | 1^{(+)}_c,J \rangle = p_{\rm cm}^{J-1}\,\Big(| 1_+ ,J\rangle + \sqrt{\frac{J}{1+J}}\,
\frac{\omega}{M}\,| 0,J \rangle \Big)\,,\;\,
 | 2^{(+)}_c ,J\rangle = \frac{p_{\rm cm}^{J+1}}{M}\,| 0 ,J\rangle \,.
\label{non-uni-def}
\end{eqnarray}
The projector matrix follows
\begin{eqnarray}
&& {\mathcal Y}^{(J P_+)}_{11,\mu \nu}(\bar q,q;w) = \sum_{k=1}^{J}
\frac{2\,J+1}{J\,(J+1)}\,k\,(k-1)\,
 Y^{(J-k)/2}_{qq}\, Y^{(J-k)/2}_{\bar q\bar q} \,Y^{k-2}_{q\bar q}
\nonumber\\
&&\quad \quad \times \Bigg(
- \frac{1}{k-1}\,c^{(J)}_{k}\,Y_{\bar q q}\,L_{1,\mu \nu}
+c^{(J)}_{k}\,L_{5,\mu \nu}
\nonumber\\
&& \quad\qquad + c^{(J-1)}_{k}\,\frac{w \cdot\bar  q}{\sqrt{w^2}}\,
\left(\frac{Y_{q q}}{Y_{\bar q \bar q}}\right)^{1/2}\,L_{3, \mu \nu}
+ c^{(J-1)}_{k}\,\frac{w \cdot q}{\sqrt{w^2}}\,
\left(\frac{Y_{\bar q \bar q}}{Y_{q q}}\right)^{1/2}\,L_{4, \mu \nu}
\nonumber\\
&& \quad\qquad +\frac{k-1-J}{k-1}\,c^{(J-1)}_{k}\,
\frac{(\bar q \cdot w)\,(w \cdot q)}{w^2}\,
\frac{Y_{q \bar q}}{(Y_{\bar q \bar q}\,Y_{q q})^{1/2}}\,L_{2,\mu \nu}
\Bigg) \,,
\nonumber\\
&& {\mathcal Y}^{(J P_+)}_{12,\mu \nu}(\bar q,q;w) = \sum_{k=1}^{J}
\frac{2\,J+1}{\sqrt{J\,(J+1)}}\,k\,
 Y^{(J-k)/2}_{qq}\, Y^{(J-k)/2}_{\bar q\bar q} \,Y^{k-1}_{q\bar q}
\nonumber\\
&& \quad\quad \times \Bigg(
-c^{(J-1)}_{k}\, \frac{\bar q \cdot w}{\sqrt{w^2}}\,
\left( \frac{Y_{q q}}{Y_{\bar q \bar q}}\right)^{1/2}\,L_{2,\mu \nu}
 - c^{(J)}_{k}\,L_{4, \mu \nu}
\Bigg) \,,
\nonumber\\
&& {\mathcal Y}^{(J P_+)}_{21,\mu \nu}(\bar q,q;w) = \sum_{k=1}^{J}
\frac{2\,J+1}{\sqrt{J\,(J+1)}}\,k\,
 Y^{(J-k)/2}_{qq}\, Y^{(J-k)/2}_{\bar q\bar q} \,Y^{k-1}_{q\bar q}
\nonumber\\
&& \quad\quad \times \Bigg(
-c^{(J-1)}_{k}\, \frac{w \cdot q}{\sqrt{w^2}}
\left( \frac{Y_{\bar q \bar q}}{Y_{q q}}\right)^{1/2}L_{2,\mu \nu}
- c^{(J)}_{k}\,L_{3, \mu \nu}
\Bigg) \,,
\nonumber\\
&& {\mathcal Y}^{(J P_+)}_{22,\mu \nu}(\bar q,q;w) = \sum_{k=0}^{J}
\big(2\,J+1\big) \, Y^{(J-k)/2}_{qq}\, Y^{(J-k)/2}_{\bar q\bar q} \,Y^{k}_{q\bar q}\;
c^{(J)}_{k}\,L_{2,\mu \nu} \,,
\nonumber\\
&& P_+ = (-1)^{J+1}\,,
\label{y-plus}
\end{eqnarray}
We observe that
the projectors ${\mathcal Y}^{(J P_+)}_{ij,\mu \nu}(\bar q,q;w)$ are regular at the kinematical
surfaces defined by $(p\cdot q)^2 =p^2\,q^2$ and
$(\bar p\cdot \bar q)^2 =\bar p^2\,\bar q^2$. Moreover we point out that none of the projectors
depends on any of the masses of initial or final states. This is an important property of the
projector since it implies
that the projectors can also be applied also in the case where initial and final particle
have a spectral distribution.

In order to complete the definition of on-shell irreducibility it remains to express the
invariant amplitude $M^{(J P)}_{ij}(\sqrt{s}\,)$ in terms of the
invariant amplitudes $F_i(\bar q,q;w)$. Some algebra leads to,
\begin{eqnarray}
&& M_{11}^{(J P_-)} =
\int_{-1}^{+1} \frac{d x }{2} \,\frac{1}{(p_{\rm cm}\,\bar p_{\rm cm})^{J}}\,
\Bigg(-P_J(x)\,F_1- {\textstyle{1-x^2\over{J\,(J+1)}}}\,P'_J(x)\,p_{\rm cm}\,\bar p_{\rm cm}\,F_5 \Bigg) \,,
\nonumber\\ \nonumber\\
&& M_{11}^{(J P_+)}=\int_{-1}^{+1} \frac{d x }{2} \,\frac{1}{(p_{\rm cm}\,\bar p_{\rm cm})^{J-1}}
\,
\Bigg(\Big( -{\textstyle{1-x^2\over{J\,(J+1)}}}\,P'_J(x)-x\,\,P_{J}(x)\Big)\,F_1
\nonumber\\
&& \qquad \qquad \quad \quad  +\Big({\textstyle{1-x^2\over{J\,(J+1)}}}\,x\,P_J'(x)- (1-x^2)\, P_J(x)\Big)\,p_{\rm cm}\,\bar p_{\rm cm}\,F_5 \Bigg) \,,
\nonumber\\
&& M_{12}^{(J P_+)}=
\int_{-1}^{+1} \frac{d x }{2} \,
\frac{\bar p_{\rm cm}^2}{(p_{\rm cm}\,\bar p_{\rm cm})^{J+1}}\,
\Bigg( \sqrt{{\textstyle{J\over{J+1}}}}\,P_{J+1}(x)\,\omega\,F_1
-{\textstyle{1-x^2\over{\sqrt{J\,(J+1)}}}}\,P_J'(x) \, p_{\rm cm}^2\,F_4
\nonumber\\
&& \qquad \qquad\quad \quad +
\sqrt{{\textstyle{J\over{J+1}}}}\,\Big( P_J(x)-x\,P_{J+1}(x)\Big)\,\omega\,p_{\rm cm}\,\bar p_{\rm cm}\,F_5\Bigg)  \,,
\nonumber\\
&& M_{21}^{(J P_+)}=
\int_{-1}^{+1} \frac{d x }{2} \,
\frac{p_{\rm cm}^2}{(p_{\rm cm}\,\bar p_{\rm cm})^{J+1}}\,
\Bigg( \sqrt{{\textstyle{J\over{J+1}}}}\,P_{J+1}(x)\,\bar \omega\,F_1
-{\textstyle{1-x^2\over{\sqrt{J\,(J+1)}}}}\,P_J'(x) \, \bar p_{\rm cm}^2\,F_3
\nonumber\\
&& \qquad \qquad \quad\quad +
\sqrt{{\textstyle{J\over{J+1}}}}\,\Big( P_J(x)-x\,P_{J+1}(x)\Big)\,\bar \omega\,p_{\rm cm}\,\bar p_{\rm cm}\,F_5\Bigg)  \,,
\nonumber\\
&& M_{22}^{(J P_+)} =
\int_{-1}^{+1} \frac{d x }{2} \,\frac{1}{(p_{\rm cm}\,\bar p_{\rm cm})^{1+J}}\,\Bigg(
- {\textstyle{2\,J+1\over{J+1}}}\,P_{J+1}(x)\,\omega\,\bar \omega\,F_1
\nonumber\\
&& \qquad \qquad\quad\quad +P_J(x)\,p_{\rm cm}\,\bar p_{\rm cm}\,F_2
 - P_{J+1}(x) \,\Big(
 \omega\,\bar p_{\rm cm}^2\,F_3+ \bar \omega\,p_{\rm cm}^2\,F_4\Big)
\nonumber\\
&& \qquad \qquad\quad \quad+\Big({\textstyle{2\,J+1\over{J+1}}}\,x\,P_{J+1}(x) - {\textstyle{J\over{J+1}}}\,P_J(x)
\Big)\, \omega\,\bar \omega\,p_{\rm cm}\,\bar p_{\rm cm}\,F_5
\Bigg) \,,
\label{m-f}
\end{eqnarray}
where $x = \cos \theta$ and
$\omega = (M^2+p_{\rm cm}^2)^{1/2}$ and $\bar \omega = (\bar M^2+\bar p_{\rm cm}^2)^{1/2}$.
In (\ref{m-f}) the invariant amplitudes $F_i(\sqrt{s}\,, x)$ are considered as a function
of $\sqrt{s}= \omega +(m^2+p_{\rm cm}^2)^{1/2}= \bar \omega +(\bar m^2+p_{\rm cm}^2)^{1/2} $
and the scattering angle $ \theta$.

The results (\ref{y-minus}, \ref{y-plus}, \ref{m-f}) specify the notion of on-shell
irreducibility. For any two-body amplitude the on-shell irreducible part can be evaluated by
first seeking a representation in terms of the Lorentz tensors $L_{i, \mu \nu}$. The required
coefficient functions in front of the projectors can be evaluated via (\ref{m-f}).
We emphasize that the concept of on-shell irreducibility smoothly carries over to the
case where initial or final states have spectral distributions rather than well defined
energy-momentum dispersions. In this case the use of the projector makes sure that
two-body unitarity is fulfilled exactly. Though the evaluation of the on-shell irreducible
part of the effective interaction $V$ will depend on approximate mass parameters of the initial
and final states the latter will not affect the unitarity condition. All ambiguities related
to the finite width of initial or final states are moved into the interaction kernel. A particular
choice of an approximate mass parameter influences only what is leading order in the kernel and
what will be treated as a correction.

\end{document}